\pdfoutput=1

\newcommand{\degr}{$^\circ$}
\newcommand{\caproman}[1]{\uppercase\expandafter{\romannumeral#1}}
\newcommand{\bs}{$\backslash$}

\newcommand{\BGsite}{http://users.physics.harvard.edu/\raisebox{-4pt}{$\tilde{\;\;}$}gottschalk}

\newcommand{\BGmail}{bernardgottschalk\,@\,gmail.com}

\newcommand{\sect}[1]{Sec.\,\ref{sec:#1}}
\newcommand{\eqn}[1]{Eq.\,(\ref{eqn:#1})}

\newcommand{\fig}[1]{Fig.\,\ref{fig:#1}}

\documentclass[11pt]{article}

\title{\bf Radiotherapy Proton Interactions in Matter}
\author{Bernard Gottschalk\thanks{Harvard University Laboratory for Particle Physics and Cosmology, 18 Hammond St., Cambridge, MA 02138, USA, \BGmail}}
\date{\today}

\usepackage{amsmath}
\usepackage{latexsym}
\usepackage{fancyhdr}
\usepackage{graphicx}

\begin{document}

\maketitle

\begin{abstract}
\noindent Radiotherapy protons interact with matter in three ways. Multiple collisions with atomic electrons cause them to lose energy and eventually stop. Multiple collisions with atomic nuclei cause them to scatter by a few degrees. Occasional hard scatters by nuclei or their constituents throw dose out to large distances from the beam. Unlike the first two processes, these hard scatters or `nuclear interactions' do not obey any simple theory, but they are rare enough to be treated as a correction. We discuss the three interactions (stopping, multiple Coulomb scattering, and hard scatters) in turn.

We begin, however, with the fundamental formula relating dose (energy deposited per unit mass) to fluence (areal density of protons) and mass stopping power (their rate of energy loss).

Concerning stopping, we define range experimentally, noting that the clinical and physics definitions are different. We present in simplified form the `Bethe-Bloch' theory of energy loss, how it leads to range-energy tables, how to interpolate these, and two simple 2-parameter approximations to the range-energy relation. Finally we discuss range straggling, the fact that protons in a monoenergetic beam stop at slightly different depths.

Concerning multiple Coulomb scattering (MCS), we present a short reading list and point out some incorrect or unhelpful aspects of the standard literature. We outline a simplified form of Moli\`ere theory, the accepted theory for protons and correct to 1\% as far as is known. We explain the Gaussian approximation to Moli\`ere theory, which suffices for almost all radiotherapy calculations, and discuss Highland's formula, vastly simpler than the full theory. We discuss scattering power, an approximation to Moli\`ere theory required by transport calculations.

We treat hard scatters by focusing on the `nuclear halo' acquired by a bundle of protons (a `pencil beam') as it stops in matter. We emphasize recent experimental results and their implications for the most efficient parameterization of the halo.

All three interactions combine in the Bragg curve, the depth-dose distribution of a monoenergetic beam stopping in water and the signature property of charged (as distinct from neutral) radiotherapy beams. For computation, the Bragg curve should be converted to an effective mass stopping power. We discuss two limiting cases.

Looking ahead, we sketch how stopping theory and MCS combine in `Fermi-Eyges' transport theory when, as is normally the case, both processes occur at once. We present three very general properties of beams spreading in a homogeneous medium, found by Preston and Koehler at the dawn of proton radiotherapy. Finally, we sketch a dose algorithm which outperforms those in current use and may ultimately prove useful.

Five appendices contain supporting material: acronyms and symbols, a review of 1D and 2D Gaussians, relativistic single particle kinfematics and finally, a few problems that occur in beam line design.

\end{abstract}

\clearpage\tableofcontents

\clearpage
\section{Introduction}
Radiotherapy protons (kinetic energy between 3 and 300\,MeV) interact with matter in three ways. They lose energy and eventually stop by multiple electromagnetic (EM) collisions with atomic electrons (and to a much smaller degree, atomic nuclei). We call this process {\em stopping}. They scatter a few degrees by multiple EM collisions with atomic nuclei (and to a smaller degree, electrons). This is called {\em multiple Coulomb scattering} (MCS). Finally, they occasionally undergo single {\em hard scatters} either by the nucleus as a whole or by constituents of the nucleus (protons, neutrons or clusters such as $\alpha$ particles). Hard scatters are also called {\em nuclear interactions}, even though they may involve either the nuclear (strong) force or the EM force.

There exist well tested and relatively simple theories for the first two processes: Bethe-Bloch theory for stopping and Moli\`ere theory for MCS. By contrast, the theory of hard scatters is complicated and largely phenomenological. We are saved by the fact that hard scatters are relatively infrequent, occurring at roughly 1\%/cm in water. Typically, only 20\% of the protons in a radiotherapy beam suffer a hard scatter before stopping. Many physics problems, such as beam line design, can be solved by considering only stopping and MCS. For others, such as predicting the dose distribution in a patient, hard scatters must eventually come in as a correction.

Those protons that escape hard scatters are called {\em primaries}. Particles emerging from a hard scatter are {\em secondaries}. These include the incident proton, other protons, neutrons, clusters (such as $\alpha$ paticles) ejected from the nucleus, and the residual nucleus itself.\footnote{~Our definition is not universal. Some Monte Carlo programs regard elastically scattered protons as primaries. The rationale for our definition will be more obvious later.} The previous paragraph says, in other words, that the dose in a homogeneous medium, or even in a heterogeneous medium, can often be computed fairly well ignoring secondaries.

{\em Absorbed dose} (or just `dose' for short) is the energy per unit mass deposited by the beam in the stopping medium. It is measured in Gray (Gy): 1\,Gy\,$\equiv$\,1\,J/Kg. Thus, dose is a mass density which varies from point to point in the patient. The common expression `dose to the patient' makes no sense, taken literally. It usually means `the maximum dose in the target volume'. This may seem like nitpicking, but careful use of language is essential to analyzing a problem and reducing it to mathematics.

The archetypal problem of proton radiotherapy physics is that of finding the dose throughout some region of interest exposed to known proton beams. The dose at any point of interest (POI) depends upon the number of particles and their propensity to lose energy. Stated quantitatively, that leads to the concepts of {\em fluence} and {\em mass stopping power} and eventually, to the fundamental formula for dose discussed in \sect{dose}.

After laying that groundwork, we cover the three processes by which fluence, mass stopping power, and eventually dose can be computed. Stopping is discussed in \sect{stopping}, MCS in \sect{MCS}, and hard scatters in \sect{hard}. These processes underly the {\em Bragg curve}, meaning the entire depth-dose distribution of a monoenergetic beam stopping in a uniform medium (often water, the simplest proxy for tissue). {\em Bragg peak} (BP) refers to the peak in the Bragg curve near end-of-range. The Bragg peak is the defining characteristic of radiotherapy protons and other charged particle beams. \sect{Bragg} covers the Bragg curve and the closely related concept of {\em effective stopping power}.

That finishes the fundamentals. Although some interesting problems involve stopping or MCS separately (Appendix \ref{sec:TypicalProblems}), to compute dose in general we must recognize that stopping and MCS happen {\em simultaneously} in the stopping medium, thus requiring a {\em transport theory}. Luckily that exists, in the form of {\em Fermi-Eyges} (FE) theory, and a comprehensive procedure for finding dose, a {\em dose algorithm}, can be built on it. FE theory is sketched in \sect{FE}. The three properties of beam spreading in homogeneous media found by Preston and Koehler, which also follow from FE theory, are given in \sect{PK}. We close by outlining a general proton dose algorithm which may someday prove useful (\sect{PBA}).

Some supporting material is contained in appendices. \ref{sec:Acronyms} and \ref{sec:Symbols} are lists of acronyms and symbols. Appendix \ref{sec:Gaussians} reviews 1D and 2D (cylindrical) Gaussians, explaining a potentially confusing factor $\sqrt{2}$ and citing a property of 2D Gaussians that limits their efficiency and therefore, the dose rate in single-scattered beams. Appendix \ref{sec:Kinematics} gives some relativistic single particle equations relating kinetic energy (KE) to speed $v$, momentum $p$ and the quantity $pv$ arising in MCS theory.\footnote{~Radiotherapy protons are somewhat relativistic: at 160\,MeV, $v\approx0.5\,c$. If we use the exact formulas we do not have to worry whether the non-relativistic limit is adequate.} Finally, Appendix \ref{sec:TypicalProblems} discusses typical problems of the sort that arise in beam line design, and introduces a program LOOKUP written to solve them.

\section{The Fundamental Formula for Dose}\label{sec:dose}
Suppose we wish to compute the dose $D(x,y,z)=D(\vec{x})$ throughout a known heterogeneous terrain (e.g. a beam-spreading  line followed by a water tank or patient) exposed to known proton beams. To start, consider a far simpler case: an infinitesimal volume of frontal area $dA$ and thickness $dz$ exposed to $dN$ monoenergetic protons at normal incidence. Then
\begin{equation}D\;\equiv\;\frac{\mathrm{energy}}{\mathrm{mass}}\;=\;
\frac{-\,dN\,(dT/dz)\,dz}{\rho\,dA\,dz}\;=\;\left(\frac{dN}{dA}\right)\times\left(-\frac{1}{\rho}\frac{dT}{dz}\right)\;=\;\Phi\times \frac{S}{\rho}\label{eqn:D}\end{equation} 
\begin{center}dose$\;=\;$fluence$\;\times\;$mass stopping power\end{center}
\eqn{D} is the starting point of every dose calculation.\footnote{~The minus sign makes $D$ positive because energy {\em decreases} with $z$. Also, our definition of fluence assumes the protons are all more or less normal to $dA$ as is often the case. Otherwise a more general definition is required \cite{Attix1986,Palmans2015}.} However, the terminology is often disguised. For instance, $S/\rho$ is frequently called the `central axis term' and $\Phi$ the `off axis term'. That language obscures the essential point: in a particle beam, the dose at any point equals the areal density of particles times their rate, with $z$, of energy loss.

\eqn{D} is inconvenient in MKS units. Also, it is easier to use total proton {\em charge} rather than the number of protons. In practical units it follows that
\begin{equation}D\;=\;\Phi_q\;(S/\rho)\;=\;(q/A)\;(S/\rho)\quad\hbox{Gy}\label{eqn:Dq}\end{equation}
with $q/A$ in nC/cm$^2$ and $S/\rho$ in MeV/(g/cm$^2$), as it is usually tabulated. Finally, if we are interested in dose {\em rate} we take the derivative with respect to time obtaining
\begin{equation}\dot{D}\;=\;\dot\Phi_q\;(S/\rho)\;=\;(i/A)\;(S/\rho)\quad\hbox{Gy/s}\label{eqn:Di}\end{equation}
with $i/A$ in nA/cm$^2$.

To illustrate, the dose rate at the surface of a water cylinder of radius 10\,cm irradiated uniformly by 1\,nA of 160\,MeV protons is (anticipating \sect{CSDA} for $S/\rho$)
\[\frac{1\;\mathrm{nA}}{\pi\;10^2\;\hbox{cm$^2$}}\times5.167\;\frac{\hbox{MeV}}{\hbox{g/cm$^2$}}\;=\;
  0.0164\;\frac{\hbox{Gy}}{\hbox{s}}\;\approx\; 1\;\frac{\hbox{Gy}}{\hbox{minute}}\] 
Since 1\,Gy/min is a reasonable clinical dose rate, we have already learned something useful: proton radiotherapy currents are on the order of nA.

In more complicated cases, multiple proton bundles (`pencil beams') of different energies may pass near enough to the POI to contribute some dose. Fluence, stopping power and dose at the POI must then be calculated separately for each pencil beam, and the doses added.

If `dose' is not qualified, it means dose to the material at the POI. That is often not what we want to know. For instance, in a dosimeter calibration beam the dosimeter might actually be in air. However, we are not interested in the dose to air at the POI, or even the dose to the materials making up the dosimeter, but rather the dose that would be delivered to a small imaginary volume of water ({\em dose to water}) at that same point. In that case we will compute $\Phi$ and proton energy at the POI assuming the actual beam line including air and a geometric model of the dosimeter, but we will use $S/\rho$ for water at that energy.

\section{Stopping}\label{sec:stopping}
Most radiotherapy protons traversing matter slow down and eventually stop by myriad soft EM collisions with atomic electrons. (These are the `primaries' defined earlier.) The present section concerns the {\em continuous slowing down approximation} (CSDA) theory of this process, range-energy tables, and some consequences for proton radiotherapy calculations.

\subsection{Experimental Definition of Range}
\begin{figure}[p]
\centering\includegraphics[width=6in,height=2.4in]{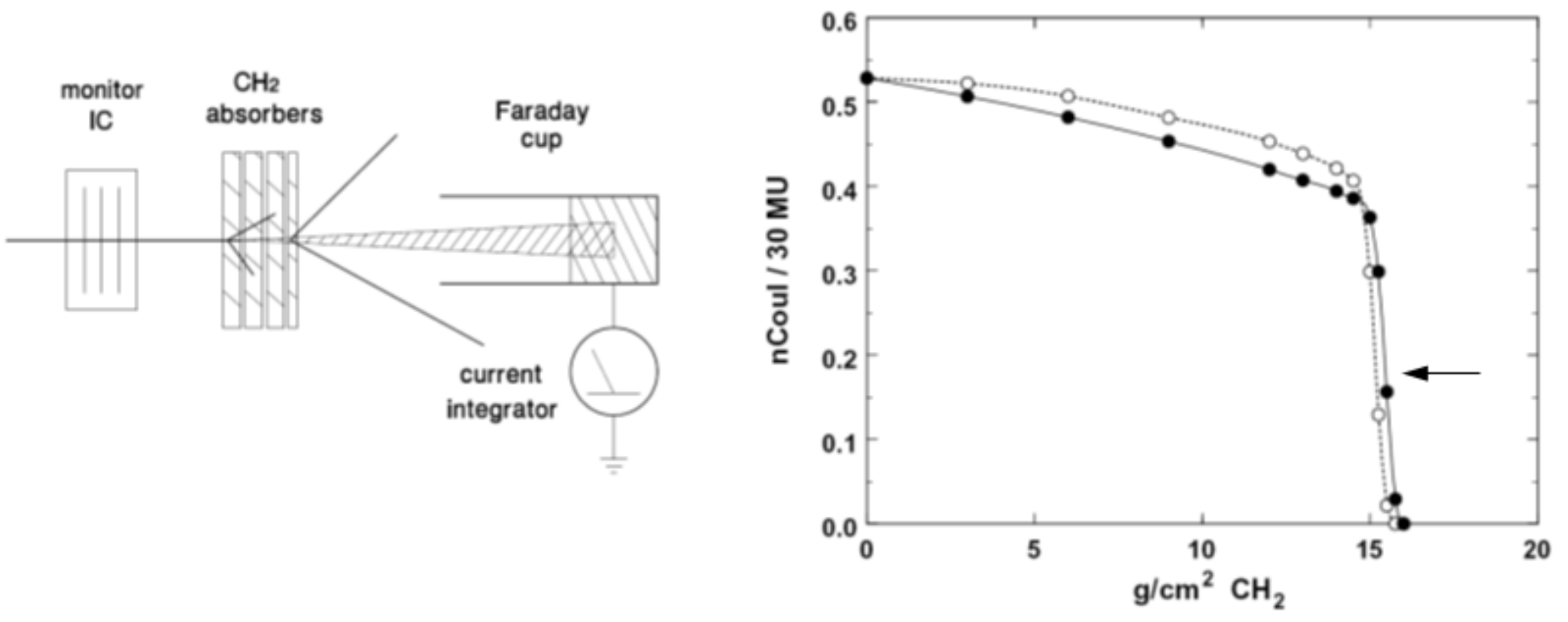}
\caption{Measuring range with a Faraday cup (FC). Filled circles correspond to the setup shown; open circles, to a FC placed much closer to the material under test. The arrow points to the mean projected range.}\label{fig:FCexperiment}
  \end{figure}

The {\em mean projected range} of primaries (which we will simply call {\em range}) can be measured as shown in \fig{FCexperiment}. A beam monitor is followed by a variable-thockness column of the material under test. The protons stop in a Faraday cup (FC) which measures total charge, acting as a proton counter. In the geometry shown (FC far from the stack) the FC catches all the primaries but few secondaries, because of their large angles and low energy. The measured charge (filled circles) at first falls slowly and almost linearly with stack thickness at $\approx1\%$/(g/cm$^2$) as primaries are lost to hard scatters. It then falls steeply to zero as the primaries range out. The range $R$ (cm) or mass range $\rho R$ (g/cm$^2$) of the beam is defined as the halfway point of the steep part. It is the depth at which half the surviving primaries (n.b. not half the incident protons!) have stopped.

If the FC shown schematically in \fig{FCexperiment} (see  \cite{verhey} for a full description) is replaced by a vacuumless or `poor man's' Faraday cup (PMFC) \cite{CascioFC} placed much nearer the stack, the curve changes to that shown in open circles because some secondaries now enter the FC.\footnote{~The sharp dropoff shifts slightly because the entrance window of this PMFC happened to be thicker.}

\begin{figure}[p]
\centering\includegraphics[width=2.68in,height=4in]{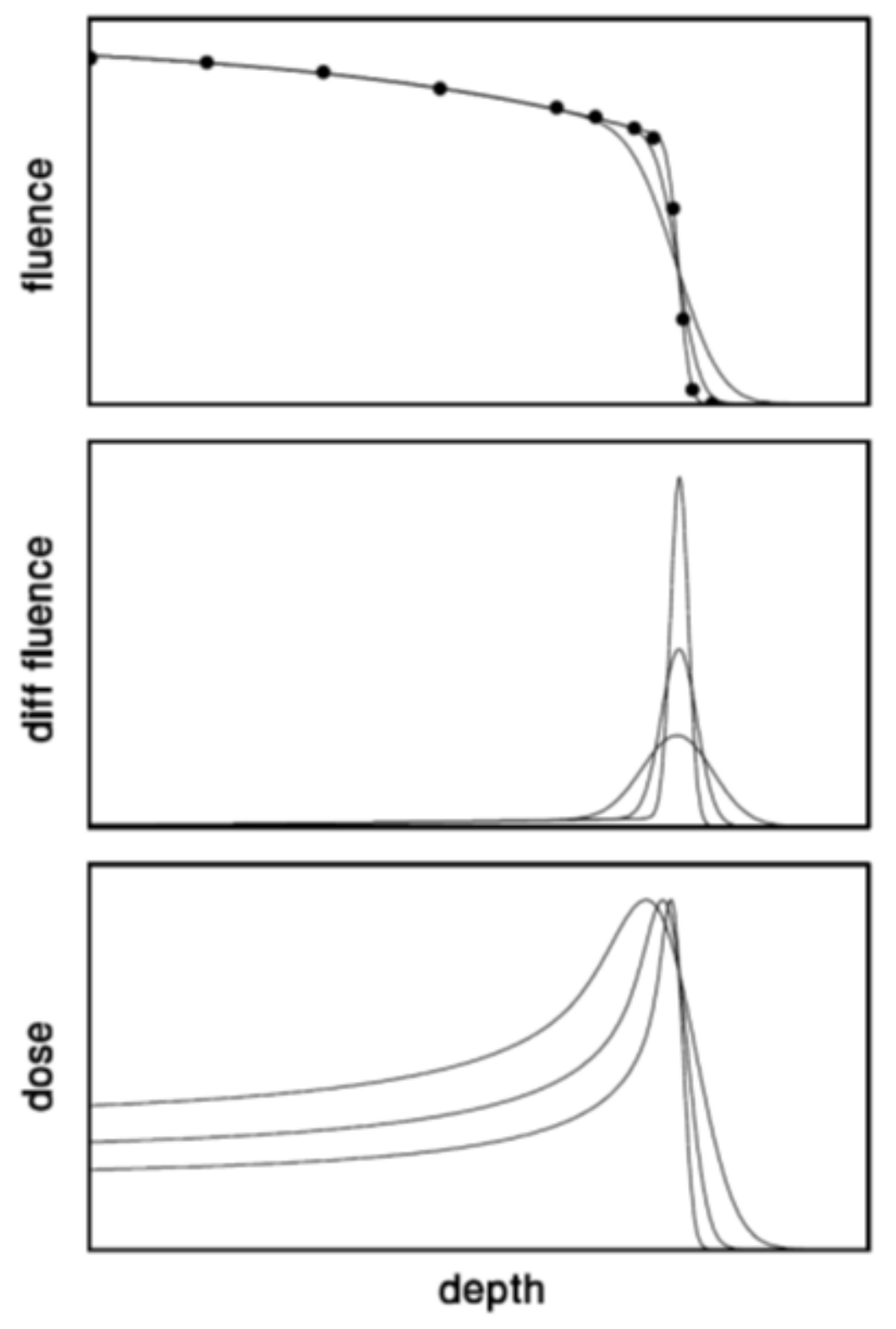}
\caption{Illustrating $R=d_{80}$. Top panel: range measured with a simple FC. Middle panel: range measured with a multilayer FC. Bottom panel: range measured with a dosimeter in a water tank. Beams with the same mean energy but different energy spreads are shown. $R$ is invariant in the FC curves whereas $d_{80}\,(=R)$ is invariant in the Bragg curve.\label{fig:d80}}
  \end{figure}

Alternatively one can use a multilayer Faraday cup (MLFC) (see \cite{BGMLFC,mlfc1} for descriptions) obtaining the middle panel of \fig{d80}, where the range corresponds to the peak channel, or more precisely, the mean channel of the sharp peak. The MLFC allows the range to be verified to a fraction of a millimeter (water equivalent) in a matter of seconds \cite{rvcal}.

In clinical facilities, range is frequently measured with a dosimeter and a water tank yielding a Bragg curve, which is dose v. depth, {\em not} number of protons v. depth. That raises a question: what point on the Bragg curve corresponds to the mean projected range just defined in terms of FC measurements? The answer
\begin{equation}\label{eqn:d80}
R\;=\;d_{80}
\end{equation}
that is, the mean projected range equals the depth of the distal 80\% point of the Bragg peak, was first found by A.M. Koehler \cite{amkpriv}. It has been confirmed many times since then e.g.\cite{berger3,bortfeld}. It follows from the fundamental equation \eqn{D} as applied to a depth-dose measurement in water.\footnote{~If we want the range of protons entering the proton nozzle we must of course correct for the range loss in scatterers, air, tank wall, and dosimeter wall.}\footnote{~In particle beams depth-dose measurements are best made with a plane-parallel ionization chamber (IC) rather than a cylindrical IC \cite{bichselic}.}

As Figure\,\ref{fig:d80} shows, if we increase the energy spread of the proton beam, keeping the mean energy the same, the mean projected range measured all three ways, in particular using \eqn{d80}, remains the same.

In the general physics literature, such as range-energy tables, `range' means precisely the quantity defined above, namely the depth at which half the surviving primaries have stopped. Unfortunately, in clinical practice `range' is used rather loosely, often denoting $d_{90}$ rather than $d_{80}$. Consistent use of the $d_{xx}$ notation will help reduce confusion.

\subsection{Theoretical CSDA Range}\label{sec:CSDA}
The theoretical CSDA rate of energy loss of fast charged particles in matter was found by Bethe and Bloch around 1933. Good modern accounts with many references can be found in introductions to the range-energy tables of Janni \cite{janni66,janni82} and ICRU Report\;49 \cite{icru49}. If we restrict ourselves to 3---300\,MeV protons (neither very low nor very high energy), most of the corrections described in those accounts are negligible, as is the energy loss to recoil nuclei. Taking advantage of that and applicable kinematic approximations, the mass stopping power in an elementary material of atomic number $Z$ and atomic mass $A$ is
\begin{equation}\label{eqn:bethe}
\frac{S}{\rho}\;\equiv\;-\frac{1}{\rho}\frac{dT}{dz}\;=\;0.3072\;\frac{Z}{A}\;\frac{1}{\beta^2}\;\left(\ln\frac{W_m}{I}-\beta^2\right)
\qquad\frac{\mathrm{MeV}}{\mathrm{g/cm}^2} 
\end{equation}
where $\beta\equiv v/c$ of the proton, see \eqn{betasq}, and
\begin{equation}\label{Wm}
W_m\;=\;\frac{2\,m_ec^2\,\beta^2}{1-\beta^2}\qquad,
\end{equation}
is the largest possible proton energy loss in a single collision with a free electron (rest energy $m_ec^2\approx0.511$\,MeV). $I$ is the {\em mean excitation energy} of the target material, a critical parameter to which we will return. 

Once we know $S/\rho$ as a function of $\beta$ (that is, of $T$) we can find the range of a proton by imagining that it enters the material at $T_\mathrm{initial}$ and subtracting the energy lost in very thin slabs until the energy reaches some low value $T_\mathrm{final}$ (but not 0, because Eq.\;\ref{eqn:bethe} diverges there). The choice of $T_\mathrm{final}$ is not critical. Any very small value will do. Thus
\begin{equation}\label{eqn:range}
\rho\,R\;(T_\mathrm{initial})\;=\;\int_{T_\mathrm{initial}}^{T_\mathrm{final}}\left({\frac{1}{\rho}\frac{dT}{dx}}\right)^{-1}dT\;=\;
\int_{T_\mathrm{final}}^{T_\mathrm{initial}}\frac{dT}{S/\rho\,(T)}
\end{equation}
is the theoretical CSDA range in g/cm$^2$. Because of MCS, protons actually travel a wiggly path so strictly speaking the quantity we have computed is total pathlength rather than mean projected range, $\rho R$ being smaller by a `detour factor'. However, that correction is also negligible in the clinical regime (0.9988 for 100\,MeV protons in water \cite{icru49}). Here is a short range-energy table for water \cite{janni82}:
\begin{center}\begin{tabular}{lccccccl} 
kinetic energy&1&3&10&30&100&300&MeV\\ 
range&0.002&0.015&0.125&0.896&7.793&51.87&cm H$_2$O\\ 
\end{tabular}\end{center}
justifying our choice of 3\,-\,300\,MeV as the clinical regime. 

Having outlined the computation of $\rho R$ we return to the mean excitation energy. $I$ can not be calculated accurately from first principles so it is, in effect, an adjustable parameter of the theory. It depends on target material and is found by fitting measured ranges or stopping powers, when available, and by interpolation otherwise. It is roughly proportional to $Z$ ($I\approx10\,Z$\,eV) but irregularities due to atomic shell structure make interpolation difficult \cite{icru49}. Fortunately $S/\rho$ is logarithmic in $I$ so that, at 100\,MeV for instance, a relative increase of 1\% in $I$ only causes a $\approx0.15$\% decrease in $S/\rho$.

If the stopping material is a mixture of elements the atoms act separately and we can replace the mixture by a succession of thin sheets of each constituent element. That picture leads easily to the {\em Bragg additivity rule}
\begin{equation}\label{eqn:BraggAdd}
\frac{S}{\rho}\;=\;\sum_i\,w_i\,\left(\frac{S}{\rho}\right)_i
\end{equation}
where $w_i$ is the fraction by weight of the $i^\mathrm{th}$ element. 

{\em Compounds} are more complicated since their constituent atoms are bound in molecules and do not, strictly speaking, act separately. However, the Bragg rule seems to hold quite well \cite{janni82} even then.

Water, often a proxy for tissue in radiotherapy and therefore of particular interest, is particularly complicated being a polar molecule. There is experimental evidence \cite{moyers,cascio07,Siiskonen2011} that Janni \cite{janni82}  ($\approx0.9\%$ higher) is more accurate than ICRU Report\,49 \cite{icru49} for water.

To summarize, choosing $I$ is a complicated business requiring considerable familiarity with the experimental literature. It is mainly for this reason that our practice is to obtain mass stopping power $S/\rho$ and mass range $\rho R$ values by interpolating generally accepted range-energy tables rather than by computing them {\em ab initio} as outlined above. Tables can differ from each other by 1--2\% due solely to different choices of $I$. For the same reason, a given set of tables may be better for some materials than others.

1\% of range at 180\,MeV corresponds to $\approx2$\,mm water. Therefore when the treatment depth itself depends on it we must rely on {\em measured} ranges in water and {\em measured} water equivalents of other materials, rather than range-energy tables! Even so, the tables are invaluable in many other calculations such as beam line design.

A final note on compounds. If a compound is not listed in one of the standard tables it might seem that we need to look up $S/\rho$ for each constituent, combine them using the Bragg rule, and integrate the resulting stopping power table to obtain $\rho\,R$. However a shortcut is provided by Rasouli et al. \cite{Rasouli2015} who combine ranges directly in the simple formula
\begin{equation}\label{eqn:Rasouli}
\frac{1}{\rho\,R}\;=\;\sum_i\frac{w_i}{(\rho\,R)_i}
\end{equation}
which assumes the Bragg rule holds, and is better than $\pm0.03\%$ for an arbitrary collection of 10 compounds over ranges from 0 to 45\,cm \cite{Rasouli2015}. Since Janni \cite{janni82} gives the range-energy relation for all chemical elements, \eqn{Rasouli} easily yields the proton range in any compound or mixture at any energy.

\subsection{Interpolating Range-Energy Tables}
\begin{figure}[p]
\centering\includegraphics[width=4.5in,height=3.5in]{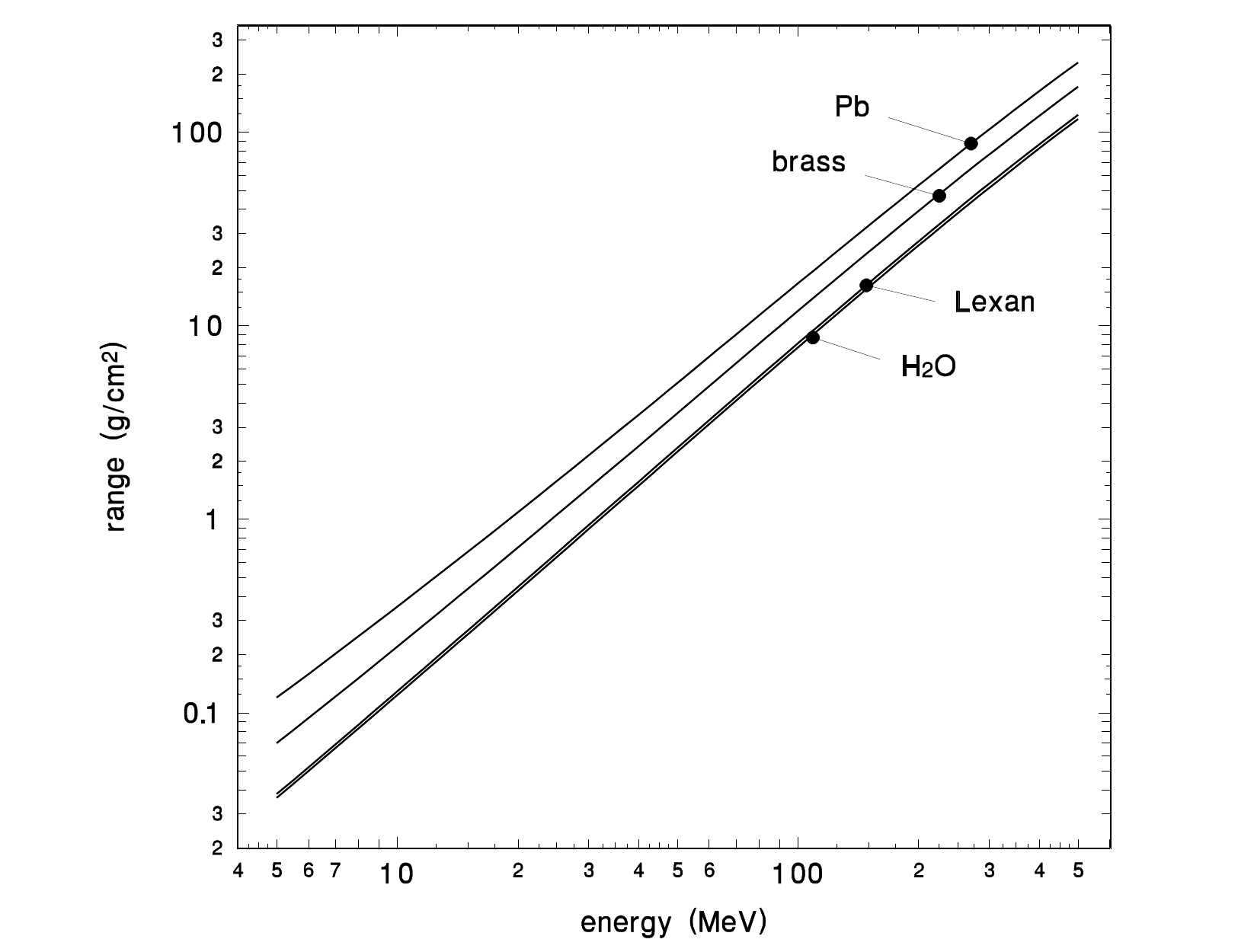}
\caption{Range-energy relation of radiotherapy protons in four useful materials \cite{icru49}.\label{fig:range}}
  \end{figure}

\fig{range} shows the range-energy relation for some useful materials in the clinical regime. At a given energy, range is greater (stopping power is lower) for heavy materials. A plot of $\log R\,(\log T)$ is nearly linear. If it were exactly linear, the relation would be an exact power law $R=aT^b$, and if the lines were parallel, $b$ would be the same for all materials. As it is, the power varies with material as well as $T$ (the lines are not quite parallel or straight). For maximum accuracy, therefore, we parameterize the log-log graph for each material separately, using cubic spline interpolation \cite{nr} of tabulated values at  1, 2, 5 ... 500, 1000\,MeV. Even this sparse data set yields $\pm0.1\%$ accuracy in the clinical regime. This is the method used by LOOKUP \cite{lookup}.

For interpolation by hand, use {\em power-law} interpolation $R=aT^b$, particularly if the table steps are large. Linear interpolation always yields too large an answer because $R(T)$ is concave upwards.

\begin{figure}[p] 
\centering\includegraphics[bb=20 20 592 465,width=4.5in,height=3.5in]{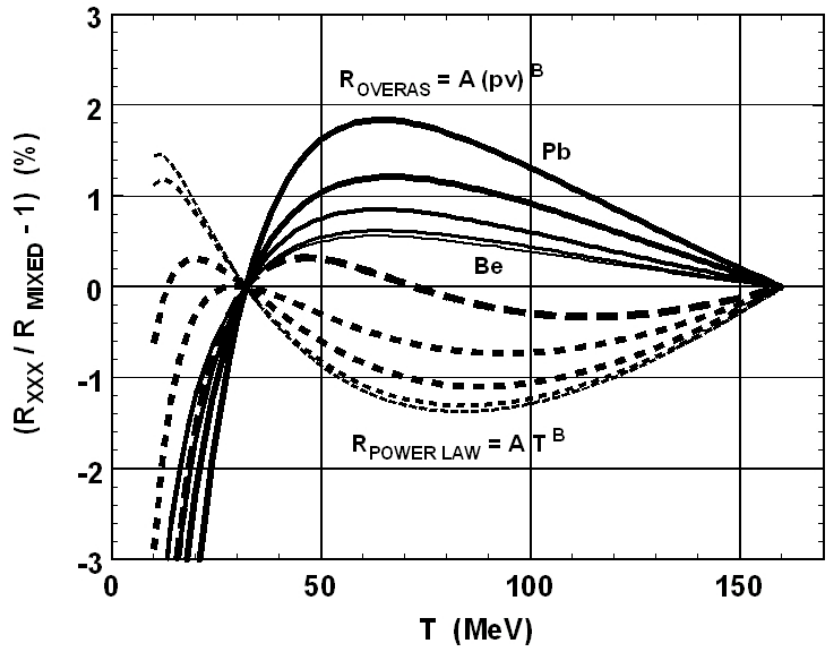}
\caption{Comparison of the {\O}ver{\aa}s range-energy approximation (solid lines) to the more common $R=aT^b$ (dashed lines). Materials in order of increasing line weight are Be, water, Al, Cu, Pb. Parameters in this example are computed from exact fits at 32 and 160\,MeV. The MIXED range-energy table corresponds to ICRU\,49 \cite{icru49} except for water which uses Janni\,82 \cite{janni82}.\label{fig:TestRover}}
  \end{figure}

When seeking answers in closed form as in \cite{bortfeld}, two-parameter approximations may be useful despite their reduced accuracy. We have already mentioned the power law $R=aT^b$. Another is the {\em {\O}ver{\aa}s approximation} \cite{overas} which can be written $R=a(pv)^b$ and takes advantage of the close relation \eqn{pv} between $pv$ and $T$.
The two approximations have comparable accuracy ($\approx2\%$) but are seen (\fig{TestRover}) to be complementary. {\O}ver{\aa}s is better for light materials and $R=aT^b$ for heavy materials. Details depend on the fiducial energies (here 32 and 160\,MeV), and it is possible to make one look considerably better than the other in particular cases. There are even cases \cite{preston,transport2012} where the `weak' {\O}ver{\aa}s approximation $R=A(pv)^2$ (different $A$) is sufficiently good.

\subsection{Range Straggling}\label{sec:rangeStraggling}
\begin{figure}[p]
\centering\includegraphics[width=4.75in,height=3.6in]{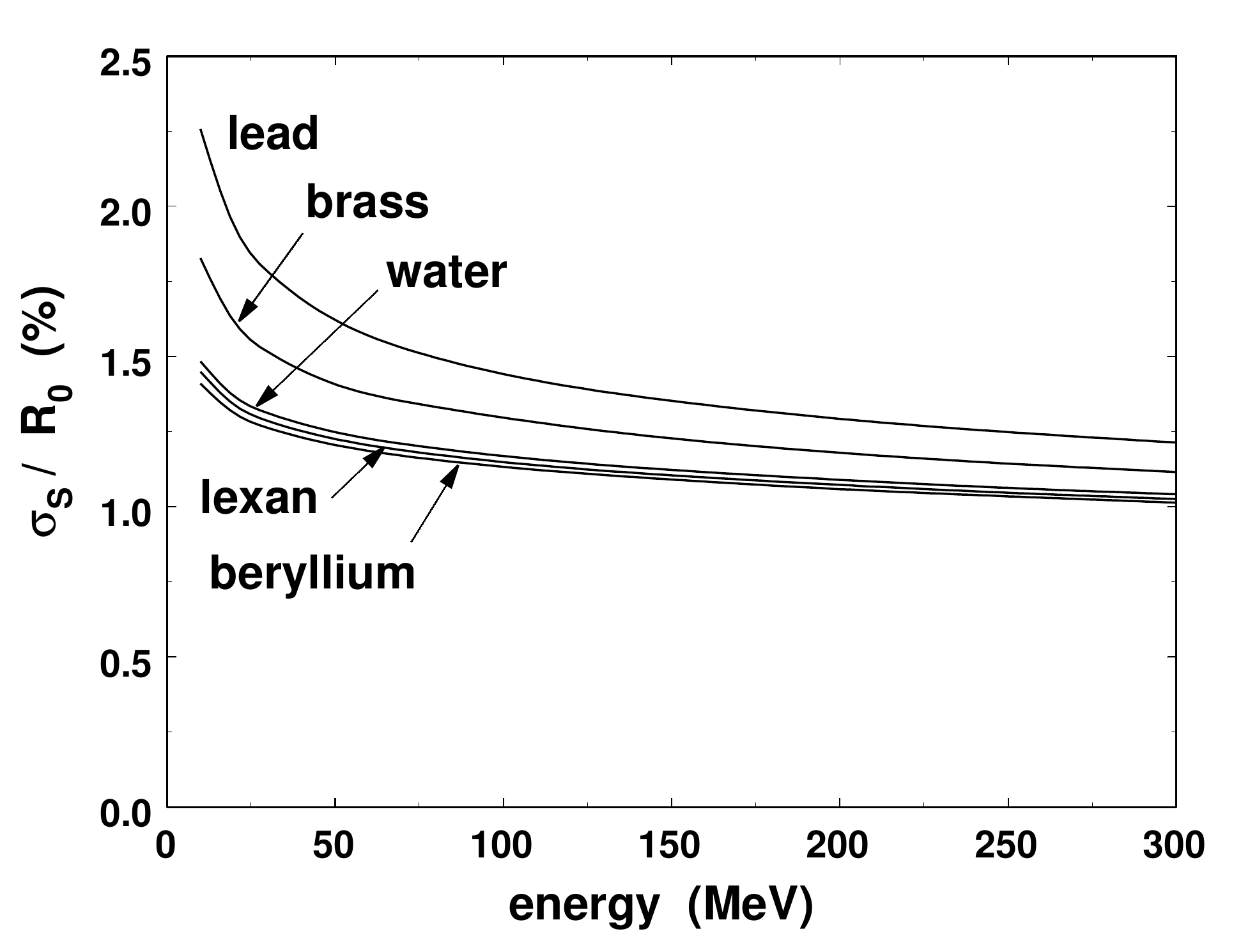}
\caption{Range straggling in several materials, from Janni \cite{janni82}.\label{fig:strag}}
  \end{figure}

Returning to \fig{FCexperiment} note that the final falloff is steep but not infinitely so. Because EM stopping involves multiple discrete and random energy transfers, protons stop at slightly different depths.  This {\em range straggling} yields a Gaussian distribution of stopping depths, characterized by $\sigma_\mathrm{S}$. See \cite{janni82} for theory and tables. \fig{strag}, a plot of Janni's $\sigma_\mathrm{S}/R$ vs. incident energy, is a summary. Roughly speaking, $\sigma_\mathrm{S}$ is a constant fraction of range. For light materials the coefficient is $\approx1.2\%$.

The fact that the coefficient is not very material-dependent is of practical importance. If materials other than water reduce the incident proton energy, the Bragg curve will not be very different from that observed in water alone. That greatly simplifies the design of passive range modulators, or the computation of dose in a heterogenous terrain.

Due to range straggling, at any depth in a water tank protons will have a distribution of energies ({\em energy straggling}) even if the incident beam is monoenergetic. The same applies to protons leaving any finite slab of any material. Most pronounced for near-stopping depths or slab thicknesses, this is just another face of the same phenomenon. The energy distribution will be still broader if the incident beam also has an energy spread.

\section{Multiple Coulomb Scattering}\label{sec:MCS}
{\em Multiple Coulomb scattering} (MCS) is the random deflection of protons by myriad EM interactions with atomic nuclei and (in low-$Z$ materials) atomic electrons. \fig{MCSdemo} illustrates a simple experiment. The transverse distribution on the {\em measuring plane} (MP) reflects the angular distribution emerging from the target, which is what MCS theory predicts.\footnote{~Strictly speaking, finding the spatial distribution at the MP is a `transport' problem. It is a trivial one if energy loss and MCS in the air are ignored. Using FE theory it is easy enough to include them.} 

\begin{figure}[p]
\centering\includegraphics[width=4.66in,height=3.5in]{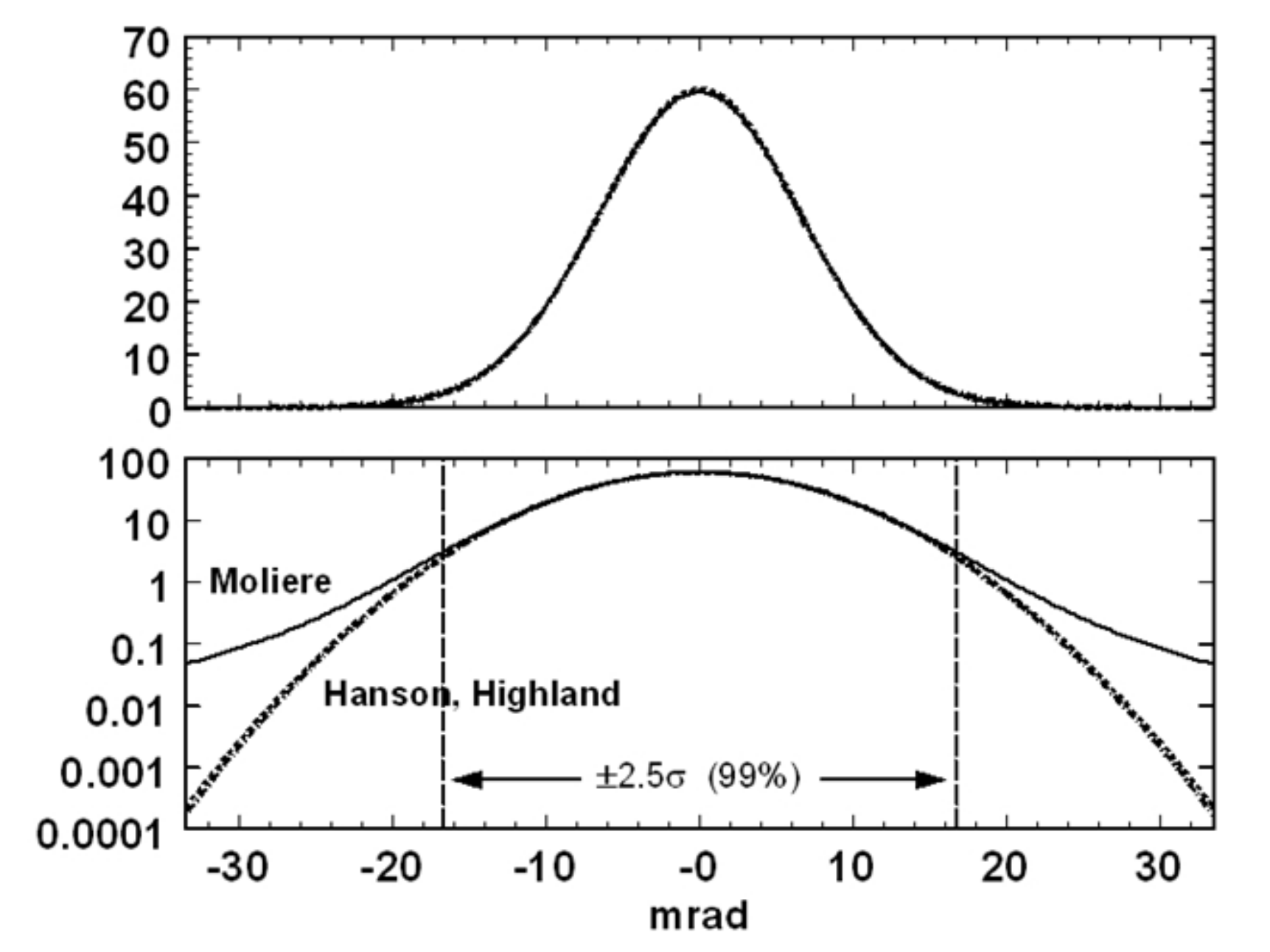}
\caption{Angular distributions for 158.6\,MeV protons traversing 1\,cm of water. On a linear plot the Moli\`ere/Fano distribution is indistinguishable from Gaussians using the Hanson or Highland $\theta_0$. On a log plot, the correct distribution peels away at $2.5\,\sigma$, and is more than $100\times$ higher at $5\,\sigma$. $2.5\,\sigma$ encompasses 99\% of the protons.\label{fig:GaussApprox}}
  \end{figure}

The angular distribution has a Gaussian core and a tail that falls much more slowly (\fig{GaussApprox}). The core contains some 99\% of the protons so the Gaussian approximation suffices for the great majority of proton radiotherapy problems. Then the only thing that remains for MCS theory to predict is the Gaussian width parameter $\theta_0$ (Appendix \ref{sec:Gaussians}) as a function of incident particle species and energy and the thickness and atomic properties of the target. We will focus on that rather than the complete theory which also predicts the angular distribution in the non-Gaussian region (the `single scattering tail').

\subsection{Suggested Reading}
The good news is that you only need one theory, that of Moli\`ere. It has no adjustable parameters and is thought \cite{bethe} to be accurate to 1\%. For a review including experimental evidence and many references but omitting derivations, see \cite{mcsbg}. For a theoretical derivation in English and a good account of competing theories see Bethe \cite{bethe}. (However,  the thick-target and compound-target aspects of Moli\`ere theory are missing from \cite{bethe}.) The German reader will find Moli\`ere's original papers \cite{moliere1,moliere2} elegant and comprehensive, except that experimental data were sparse at that early date.

The bad news is that much of the MCS literature is unhelpful and sometimes incorrect. Ignore claims that Moli\`ere didn't cover energy loss by the incident particle (he did, and the theory extends to near-stopping tagets), that the more complicated `NSW' theory is better (it isn't) or occasionally that Moli\`ere doesn't agree with experiment (it does). At one time, complicated schemes were proposed to improve Rossi's simple formula for the Gaussian approximation. These were all swept away by Highland's elegant parameterization \cite{highland}. 

Finally, even though adding mixed slabs in quadrature is incorrect in principle, it works in practice, and the alternative suggested by Lynch and Dahl \cite{lynch} is unworkable. We'll address that in connection with scattering power and transport theory.

\subsection{Elements of Moli\`ere Theory}
Moli\`ere theory is algebraically complicated.\footnote{~LOOKUP \cite{lookup} computes MCS theory in all its variants for any target material and thickness.} Here we will only treat the simplest case, a target consisting of a single element (atomic weight $A$, atomic number $Z$) sufficiently thin that the proton (charge number $z$, momentum $p$, speed $v$) loses negligible energy. We assume large $Z$ so that scattering by atomic electrons is negligible. We wish to compute the distribution of proton space angle $\theta$ given protons of known energy entering a target of thickness $t$ g/cm$^2$ where $t$ (for now) is much less than the proton mass range $\rho R$. 

We first calculate a {\em characteristic single scattering angle} $\chi_c$ given by 
\begin{equation}\label{eqn:chicsq}
\chi_c^2\;=\;c_3\;t/(pv)^2
\end{equation}
where
\begin{equation}\label{eqn:c3}
c_3\;\equiv\;4\;\pi N_A\Bigl(\frac{e^2}{\hbar c}\Bigr)^2(\hbar c)^2\;\frac{z^2Z^2}{A}
\end{equation}
$N_A\approx6.022\times10^{23}$\,gmolwt$^{-1}$ is Avogadro's number, $(e^2/\hbar c)\approx1/137$ is the fine structure constant and $(\hbar c)\approx197\times10^{-13}$\,MeV\,cm is the usual conversion factor. The physical interpretation of $\chi_c$ is that, on average, a proton suffers exactly one single scatter greater than $\chi_c$ in traversing the target.

Next we compute a {\em screening angle} $\chi_a$ using 
\begin{equation}\label{eqn:chiasq}
\chi_a^2\;=\;\chi_0^2\;(1.13\;+\;3.76\;\alpha^2)
\end{equation}
where
\begin{equation}\label{eqn:chi0sq}
\chi_0^2\;=\;c_2/(pc)^2
\end{equation}
and the {\em Born parameter} $\alpha$ is given by
\begin{equation}\label{eqn:alphasq}
\alpha^2\;=\;c_1/\beta^2
\end{equation}
where $\beta\equiv v/c$ of the proton, see \eqn{betasq}. The constants are
\begin{equation}\label{eqn:c1}
c_1\;\equiv\;\Bigl[\Bigl(\frac{e^2}{\hbar c}\Bigr)\;z\;Z\Bigr]^2
\end{equation}
and
\begin{equation}\label{eqn:c2}
c_2\;\equiv\;\Bigl[\frac{1}{0.885}\;\Bigl(\frac{e^2}{\hbar c}\Bigr)\;(m_ec^2)\;Z^{1/3}\Bigr]^2
\end{equation}
The screening angle is that (very small) angle at which the single scattering cross section levels off (departs from Rutherford's
$1/\theta^4$ law) because of the screening of the nuclear charge by atomic electrons. One of Moli\`ere's insights was that, though MCS depends on that angle, it is insensitive to the exact shape of the single scattering cross section near that angle.

Next we compute a quantity
\begin{equation}\label{eqn:b}
b\;=\;\ln\,\Bigl(\frac{\chi_c^2}{1.167\;\chi_a^2}\Bigr)
\end{equation}
which is the natural logarithm of the effective number of collisions in the target. Next, the {\em reduced target thickness} $B$ is defined as the root of the equation
\begin{equation}\label{eqn:bb}
B\;-\;\ln B\;=\;b
\end{equation}
which can be solved by standard numerical methods. $B$ is almost proportional to $b$ in the region of interest. 

Finally, Moli\`ere's {\em characteristic multiple scattering angle}
\begin{equation}\label{eqn:thetam}
\theta_M\;=\;\frac{1}{\sqrt{2}}\;(\chi_c\;\sqrt{B})
\end{equation}
is analogous to $\theta_0$ in the Gaussian approximation. Typically it is about 6\% larger.\footnote{~The $1/\sqrt{2}$ is ours. It makes $\theta_M$ more or less equivalent to $\theta_0$ in the Gaussian approximation.} Moli\`ere is now positioned to compute the distribution of $\theta$. Defining a reduced angle
\begin{equation}\label{eqn:theta'}
\theta'\;\equiv\;\frac{\theta}{\chi_c\sqrt{B}}
\end{equation}
he approximates the desired distribution function $f(\theta)$ by a power series in $1/B$
\begin{equation}\label{eqn:foftheta}
f(\theta)\;=\;\frac{1}{2\pi\;\theta_M^2}\;\frac{1}{2}\Bigl[f^{(0)}(\theta')+\frac{f^{(1)}(\theta')}{B}+
\frac{f^{(2)}(\theta')}{B^2}\Bigr]
\end{equation}
where
\begin{equation}\label{eqn:fn}
f^{(n)}(\theta')\;=\;\frac{1}{n!}\;\int_0^\infty y\;dy\;J_0(\theta'y)\;e^{\;y^2/4}\;\Bigl(\frac{y^2}{4}\ln\frac{y^2}{4}\Bigr)^n
\end{equation}
$f^{(0)}$ is a Gaussian
\begin{equation}\label{eqn:f0}
f^{(0)}(\theta')\;=\;2\;e^{\displaystyle\;-\;\theta'^2}
\end{equation}
Moli\`ere \cite{moliere2} gives further formulas and tables for $f^{(1)}$ and $f^{(2)}$.

The foregoing equations, with Bethe's improved tables \cite{bethe} for $f^{(1)}$ and $f^{(2)}$, permit one to evaluate the scattering probability density $f(\theta)$ if the target consists of a single chemical element with $Z\gg1$ and the energy loss is small. Except for rearrangements of physical constants to conform to modern usage, and the normalization of $f(\theta)$, our equations are identical to Moli\`ere's. We reiterate, however, that Moli\`ere \cite{moliere2} generalized them to {\em arbitrary energy loss} and to {\em compounds and mixtures} \cite{mcsbg}.

The generalization to low-$Z$ elements, where scattering by atomic electrons (not just the screened nucleus) is appreciable, is handled two ways in the literature. Bethe's approach \cite{bethe} is simply to substitute $Z(Z+1)$ for Moli\`ere's $Z^2$ wherever it appears. We call this Moli\`ere/Bethe. Fano's approach \cite{mcsbg,fano,scott} is more complicated. He computes a correction to $b$. Moli\`ere/Fano theory fits experimental proton data from 1\,MeV to 200\,Gev for a wide variety of materials and thicknesses at the few percent level \cite{mcsbg}. Since this is comparable to experimental error, we do not really know how good Moli\`ere theory is, but Bethe \cite{bethe} claims 1\%. There are no adjustable parameters!

\begin{figure}[p]
\centering\includegraphics[width=3.71in,height=3.5in]{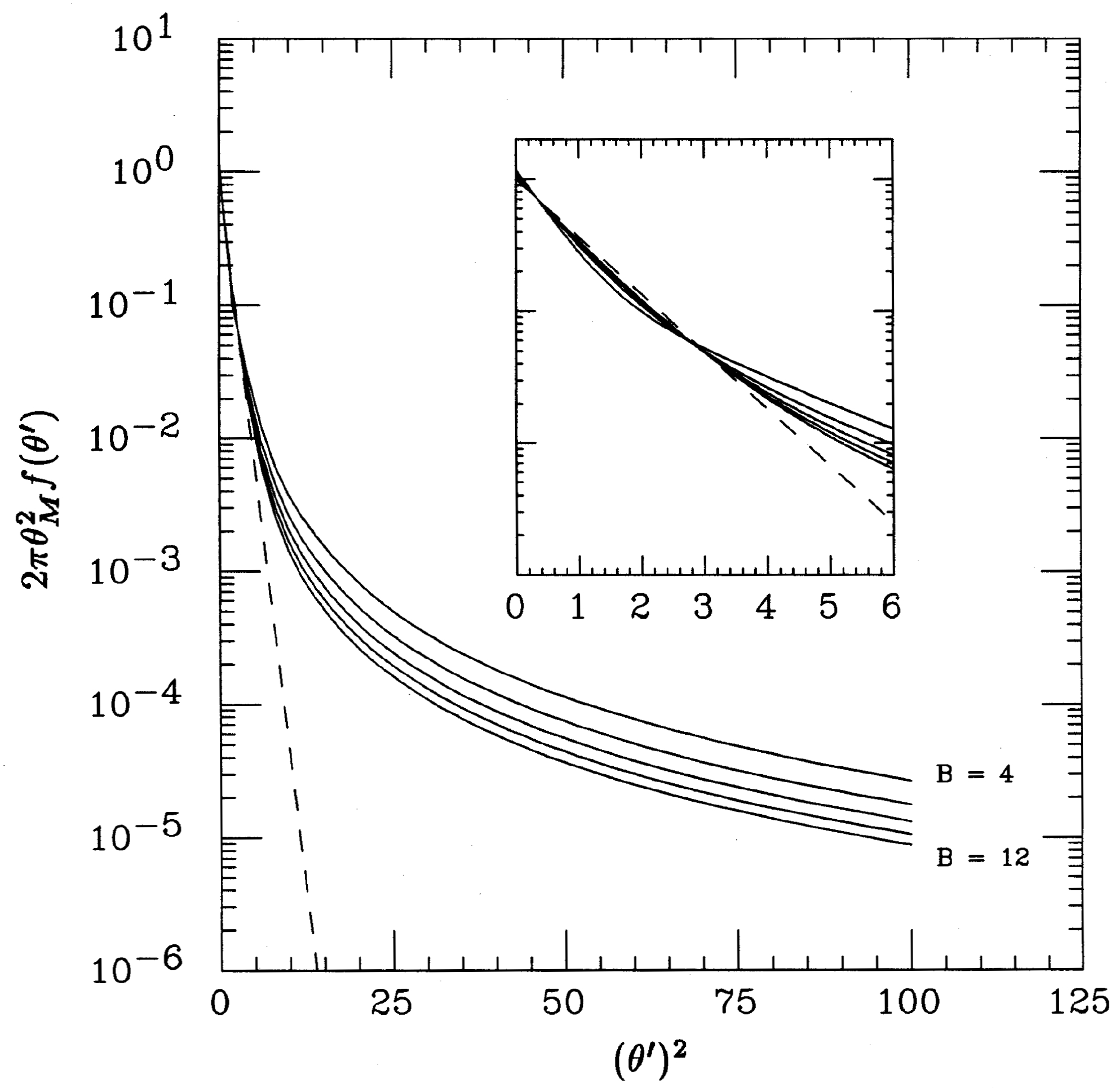}
\caption{Moli\`ere angular distribution plotted so that a Gaussian becomes a straight line \cite{mcsbg}. Dashed line: $f^{(0)}$ only. Inset: graph near the origin showing that the best-fit Gaussian is narrower than simply using $f^{(0)}$.\label{fig:mcs2}}
  \end{figure}

Though the derivation of $B$ is complicated, numerically $B$ has a simple interpretation. It is proportional to the logarithm of normalized target thickness $t/(\rho\,R)$ with a coefficient depending on the material. \fig{mcs2} shows that the angular distribution depends very weakly on $B$.

\subsection{The Gaussian Approximation}
The inset to \fig{mcs2} shows that the best-fit Gaussian is somewhat narrower than that obtained by simply dropping $f^{(1)}$ and $f^{(2)}$. This was first pointed out by Hanson et al. \cite{hanson} who, in the course of confirming Moli\`ere theory with electrons\footnote{~Bichsel \cite{bichsel} later confirmed it with protons.}, suggested using a Gaussian with a characteristic angle which we will call $\theta_\mathrm{Hanson}$: 
\begin{equation}\label{eqn:Hanson}
\theta_0\;=\;\theta_\mathrm{Hanson}\;\equiv\;\frac{1}{\sqrt{2}}\;(\chi_c\;\sqrt{B-1.2})
\end{equation}
That's not a shortcut since one still has to evaluate the full Moli\`ere theory.  In 1975, however, Highland parameterized Moli\`ere/Bethe/Hanson theory and obtained
  \begin{equation}\label{eqn:Highland}
\theta_0\;=\;\theta_\mathrm{Highland}\;\equiv\;\frac{14.1\hbox{\ MeV}}{pv}\;\sqrt{\frac{t}{\rho X_0}}\;\biggl[1\;+\;\frac{1}{9}\;\log_{10}\Bigl(\;\frac{t}{\rho X_0}\;\Bigr)\biggr]\hbox{\qquad rad}
  \end{equation}
$\rho X_0$ is the mass {\em radiation length} (g/cm$^2$) of the target material, which can be found in tables or using a standard formula \cite{rpp}. \eqn{Highland} is a huge shortcut, avoiding Moli\`ere theory entirely. It also lays bare the structure of the full theory: the Gaussian width varies primarily as the square root of target thickness, but a correction factor proportional to the logarithm of target thickness is required. It is called the {\em single scattering correction} to $\theta_0$ because it stems from the influence of single scattering at larger angles.

\begin{figure}[p]
\centering\includegraphics[width=4.48in,height=3.6in]{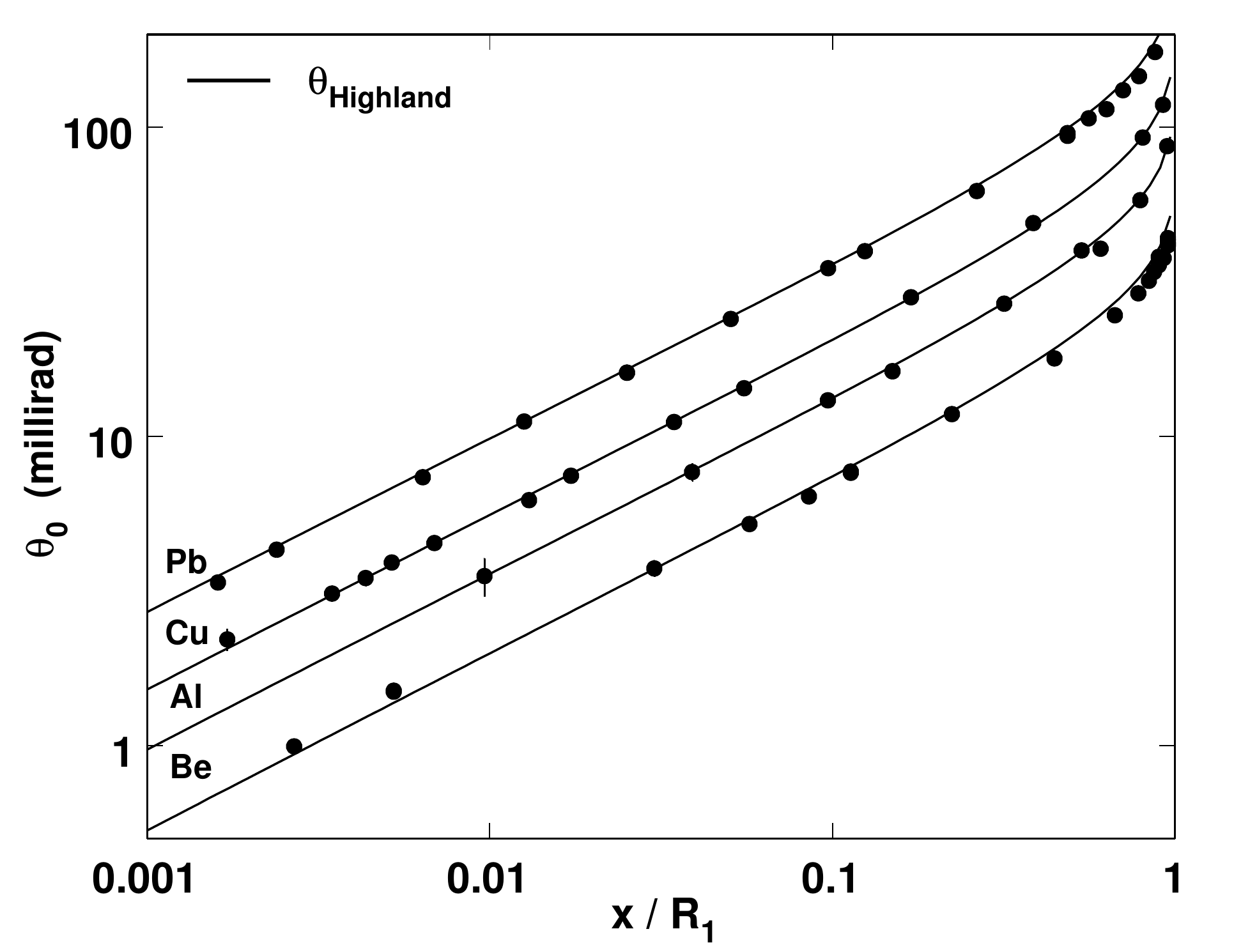}
\caption{Accuracy of Highland's formula for four elements. Abscissa is target thickness divided by proton range. Points are experimental data at 158.6\,MeV \cite{mcsbg}.\label{fig:logPlotHIG}}
  \end{figure}

\eqn{Highland} is obviously limited to thin targets, since the depth at which $pv$ is to be evaluated is not specified. (\eqn{Hanson} is {\em not} limited to thin targets.) A makeshift generalization to thick targets was proposed by us \cite{mcsbg}, namely
  \begin{equation}\label{eqn:HighlandThick}
\theta_0\;=\;14.1\,\mathrm{MeV}\,z\left[1+\frac{1}{9}\log_{10}\left(\frac{t}{\rho X_0}\right)\right]   \times\left( \int_0^t\left(\frac{1}{pv}\right)^2\frac{dt'}{\rho X_0}\right)^{1/2}
  \end{equation}
We took Highland's logarithmic correction factor out of the integral, so that it is evaluated for the {\em entire} target thickness rather than each step in the integral. (Otherwise the integral, approximated by a sum, decreases as the step size is decreased.) With that generalization, the Highland formula fits experimental data rather well (\fig{logPlotHIG}). However, \eqn{HighlandThick} has been superseded by the concept of scattering power, to which we now turn.

\subsection{Scattering Power}
Scattering power was introduced by Rossi in 1952 \cite{rossiBook} though he did not use that term. It was resurrected and named by Brahme \cite{brahme} in connection with electron radiotherapy, and now appears regularly in discussions of proton transport. This section is rather technical. It may be skipped if you are only interested in computing multiple scattering in a single homogenous slab, for which we have already given several methods. \cite{scatPower2010} is a full discussion of scattering power. What follows is just a summary.

Brahme used an analogy with stopping power, but that is imperfect. Stopping power $S(z)\equiv-\,dT/dz$ depends solely on the proton speed and atomic properties of the target  {\em at the current depth $z$} (\eqn{bethe}). It is a `local' function.  We define {\em scattering power} by    
  \begin{equation}\label{eqn:Txx}
T_\mathrm{xx}(z)\equiv d<\theta_x^2>/dz
  \end{equation}
that is, the rate of increase with $z$ of the variance of the projected MCS angle.\footnote{~The subscripts xx distinguish different formulas and differentiate scattering power from kinetic energy.} To be consistent with what we already know we must recover $\theta_\mathrm{Hanson}$ from Moli\`ere/Fano/Hanson theory if we integrate $T_\mathrm{xx}$ over the $z$ of a full slab. For that to work for any target thickness, we find \cite{scatPower2010} that $T_\mathrm{xx}$ must be {\em nonlocal}. It must depend somehow on the {\em history} of the beam as well as its energy at the POI.

Before we go into that, we consider an obvious question. Who needs scattering power, since we already know the answer? The response to that is that Moli\`ere/Fano/Hanson theory tells us only one thing: the MCS angle out of a {\em given finite thickness} of a {\em given single material}. A general transport theory (whether deterministic or Monte Carlo) needs to do much more than that. It must treat  both MCS and slowing down at the same time in mixed slabs, in a way that is independent of step size. For that we need a {\em differential} description of both processes. Stopping is easy: we already have \eqn{bethe}. But Moli\`ere theory treats only {\em finite} targets. Unlike stopping theory, it does not flow from a differential description. To answer our question: Scattering power can be viewed as a differential approximation to MCS theory, devised after the fact.  

Skipping the details, we find that six prescriptions for $T_\mathrm{xx}(z)$ occur in the literature. Some are considerably better than others in agreeing with Moli\`ere/Fano/Hanson theory when integrated over target thickness. The best and most convenient is the `differential Moli\`ere' scattering power derived by the author, namely
\begin{equation}\label{eqn:TdM}
T_\mathrm{dM}(z)\;=\;f_{dM}(pv,p_1v_1)\times\left(\frac{15.0\,\mathrm{MeV}}{pv(z)}\right)^2\;\frac{1}{X_S(z)}\quad\frac{\mathrm{rad}^2}{\mathrm{cm}}
\end{equation}
where
\begin{eqnarray}\label{eqn:fdM}
f_\mathrm{dM}&\equiv&0.5244+0.1975\log_{10}(1-(pv/p_1v_1)^2)+0.2320\log_{10}(pv/\mathrm{MeV})\nonumber\\
&&-\;0.0098\log_{10}(pv/\mathrm{MeV})\log_{10}(1-(pv/p_1v_1)^2)
\end{eqnarray}
and
\begin{equation}\label{eqn:XS}
\frac{1}{\rho\,X_S}\;\equiv\;(0.34896\times10^{-3})\times\frac{Z^2}{A}\left\{2\ln\,(33219\,(AZ)^{-1/3})-1\right\}\quad\frac{1\;}{\mathrm{g/cm}^2}
\end{equation}
This reproduces Moli\`ere/Fano/Hanson theory to $\approx\pm2\%$ for normalized target thicknesses from 0.001 to 0.97 (very thin to nearly stopping) over the full periodic table \cite{scatPower2010}.
 
$X_S$ is a {\em scattering length}, analogous to radiation length $X_0$ and of the same order of magnitude. ($X_S/X_0=1.42$ for Be decreasing monotonically to 1.04 for Pb.) In compounds and mixtures $X_S$, like $X_0$, obeys a Bragg additivity rule
  \begin{equation}
\frac{1}{\rho X_S}\;=\;\sum_{i}\frac{w_i}{(\rho X_S)_i}
  \end{equation}
$pv$ is the familiar kinematic quantity (Eq.\;\ref{eqn:pv}) at $z$, the point of interest, while $p_1v_1$ is the initial value of the same quantity. Eqs.\;\ref{eqn:TdM}\,--\,\ref{eqn:XS} apply equally well to mixed homogeneous slabs. $pv(z)$ is computed using the appropriate range-energy relation in each slab, whereas $X_S$ is a piecewise constant function of $z$.

\begin{figure}[p]
\centering\includegraphics[width=4.644in,height=3.6in]{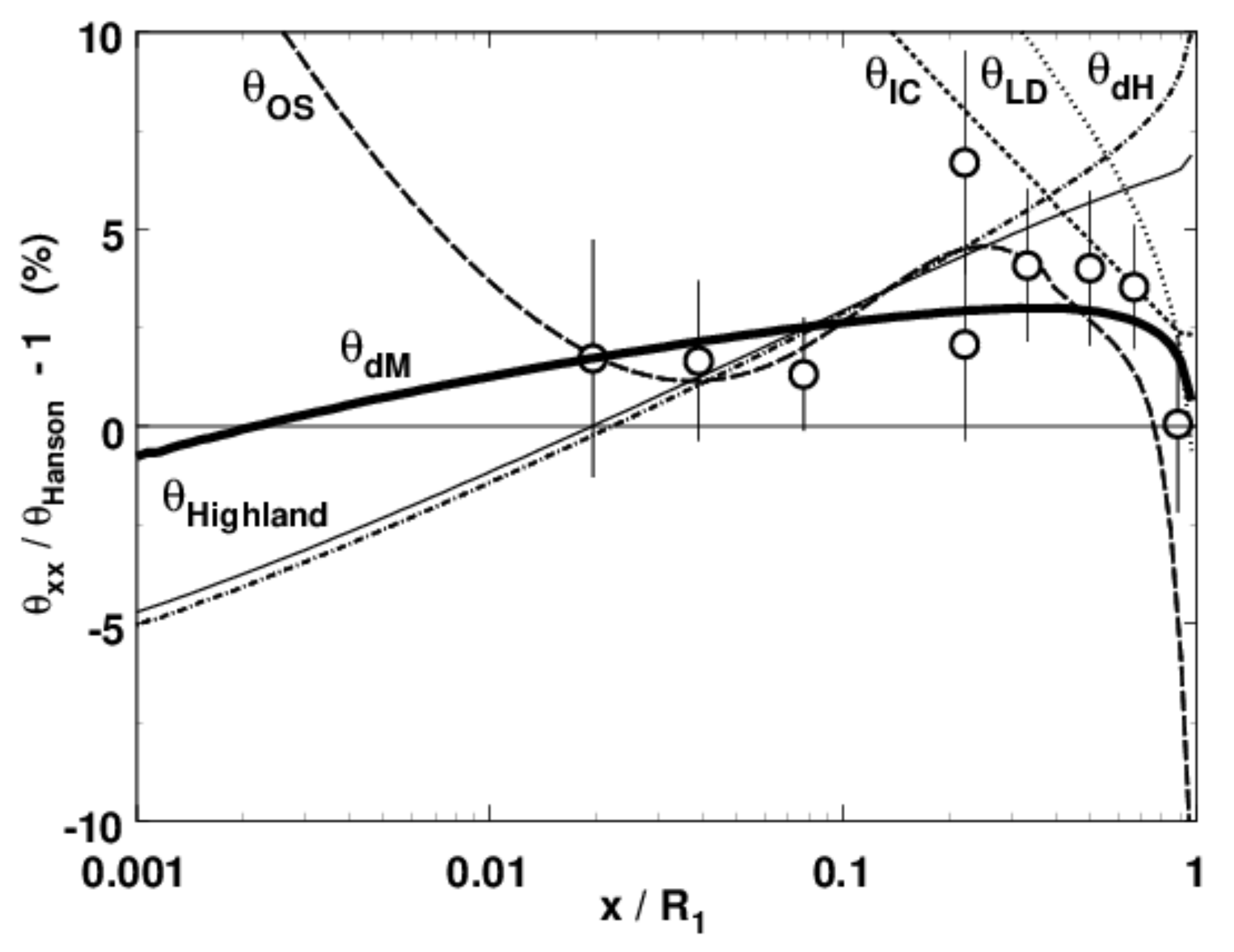}
\caption{Compliance of experiment (points, \cite{mcsbg}) and integrated $T_\mathrm{xx}$ (lines, \cite{scatPower2010}) with Moli\`ere/Fano/Hanson theory for polystyrene. The bold line represents $T_\mathrm{dM}$.\label{fig:ExptFigPoly}}
  \end{figure}

\begin{figure}[p]
\centering\includegraphics[width=4.644in,height=3.6in]{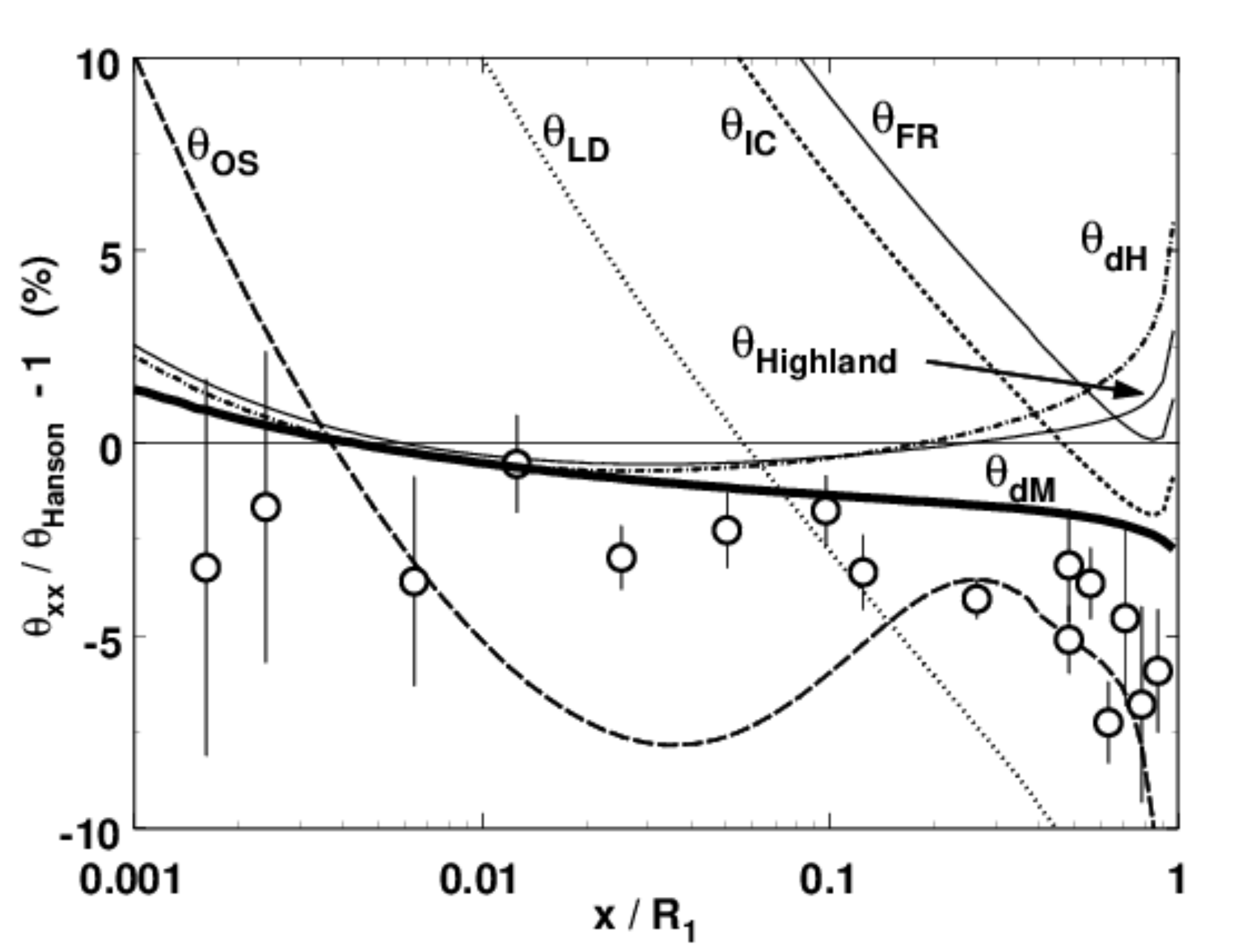}
\caption{The same as \fig{ExptFigPoly} for lead (Pb).\label{fig:ExptFigLead}}
  \end{figure}

\fig{ExptFigPoly} and \fig{ExptFigLead} demonstrate the accuracy of integrated $T_\mathrm{dM}$ for polystyrene and lead (Pb) respectively.\footnote{~Among inexpensive plastics, polystyrene is the best proxy for water. Thin-walled water targets are difficult to make.} 

An important characteristic of $T_\mathrm{dM}(z)$ is that it is {\em nonlocal} via the correction factor $f_\mathrm{dM}$. Scattering power depends not only on conditions  (that is, $pv$ and $X_S$) at $z$ but on how the protons started out ($p_1v_1$). To take a simple example, for a 20\,MeV proton in Be $d<\theta_x^2>/dz$ is smaller if the overlying thickness of Be is 0.1\,cm (protons enter at 23.7\,MeV) than if it is 5\,cm (protons enter at 102\,MeV). How can that be? How does the proton `know' what has gone before? 

The answer is that, unlike stopping and single scattering, {\em multiple Coulomb scattering is not a primitive process}. It makes sense to speak of stopping and single scattering even in an atomic monolayer, and we would expect those to depend only on proton energy and atomic properties of the monolayer. By contrast, {\em multiple} scattering of a beam is a statistical statement about many protons each undergoing many collisions, and we should not be too surprised if, in a given slab, that depends on what came before. 

The factor $f_\mathrm{dM}$ measures the beam's progress towards `Gaussianity'.  Any $T_\mathrm{xx}(z)$ that has some nonlocality built into it, some sense of the beam's history, is more accurate than any $T_\mathrm{xx}(z)$ that does not \cite{scatPower2010}. $pv$ is a particularly convenient proxy for `history' because a transport program needs to keep track of it (or kinetic energy $T$) anyway. 

A final comment on step size. We mentioned that any computation, deterministic or Monte Carlo, should converge as a function of $\Delta z$. Over some reasonable range of $\Delta z$ the answer should remain the same. $T_\mathrm{dM}$ has this property, but some Monte Carlo MCS models do not \cite{Matysiak2013}, and using a nonlocal scattering power might remedy that.

\clearpage
\section{Nuclear Interactions (Hard Scatters)}\label{sec:hard}
This section marks the greatest departure from our previous treatment \cite{paganettiBook}. There, we focused on the categorization of nuclear interactions and their relative probabilities. All that is still correct, but here we focus instead on the effect of hard scatters on the dose distribution of a pencil beam, more directly relevant to pencil beam scanning (PBS).

In addition to multiple soft EM interactions, protons stopping in matter undergo single hard scatters. These  throw dose out to large radii, creating what Pedroni et al. \cite{pedroniPencil} called the {\em nuclear halo}. A more comprehensive terminology for the anatomy of the PB dose was proposed later by us \cite{Gottschalk2015}. The {\em core} is the compact central region due to primaries, the {\em halo} is the surrounding dose from charged secondaries, and the {\em aura} is the very large region due to neutral secondaries. The three regions overlap, but each has distinct characteristics as we will see.

There may also be {\em spray}, additional low dose outside the core which enters with the beam. Unlike the halo and aura, it depends on beam line design and can in principle be avoided. It may arise from hard scatters in upstream degraders \cite{pedroniPencil}, MCS in upstream beam profile monitors \cite{Sawakuchi2010}, or protons scraping the beam pipe \cite{Lin2014}. We will ignore it here. Obviously that is not possible in practice. Spray, if present, complicates parameterization of the PB.

The halo and aura comprise a significant fraction of the total proton energy (integrated dose),  about $15\%$ at 180\,MeV. To see what that implies, suppose we build up a broad dose using pencil beam scanning, starting at the center and working outwards. As we drop in PBs nearby, their halos overlap the central PB, increasing the dose slightly. That continues until the new PBs are so far out that their halos no longer contribute. Thus, the dose at the center increases by an amount depending on field size. The consequent need to parameterize the halo as well as the core was first pointed out by Pedroni et al. \cite{pedroniPencil}.

\subsection{Contributing Reactions}
\begin{figure}[p]
\centering\includegraphics[width=4.97in,height=3.5in]{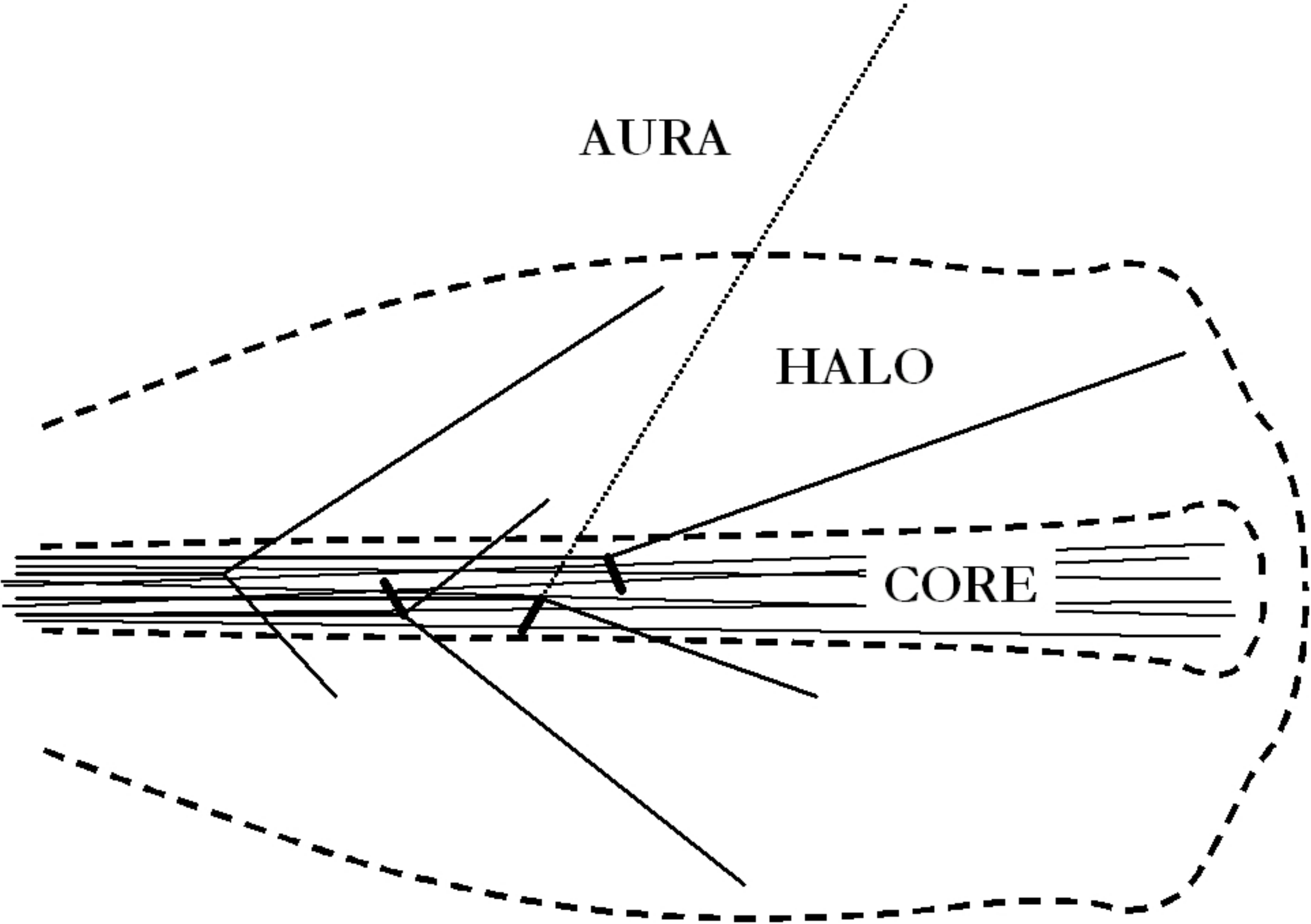}
\caption{Core, halo and aura with reactions $^1$H(p,p)p, $^{16}$O(p,2p)$^{15}$N, $^{16}$O(p,pn)$^{15}$O and $^{16}$O(p,p)$^{16}$O (from left). Recoil nuclei ranges are exaggerated. The dashed lines are 10\% and 0.01\% isodoses drawn to scale.}\label{fig:haloReactions}
  \end{figure}

Reactions leading to hard scatters can categorized in complementary ways. First of all, they may involve the EM or the nuclear force. Next, they may be {\em coherent} (interaction with the nucleus as a whole) or {\em incoherent} (interaction with a component of the nucleus). Finally, they may be {\em elastic} (kinetic energy conserved), {\em inelastic} (KE not conserved, recoil nuclide the same as the target, but excited) or {\em nonelastic} (recoil nuclide different, and possibly excited). We discuss a few such reactions, shown schematically in \fig{haloReactions}.

Hard scattering on free hydrogen  $^1$H(p,p)p is elastic\footnote{~The reaction  $^1$H(p,p$\gamma$)p is permitted but is negligible at radiotherapy energies \cite{shlaer}.}, resembling a billiard-ball collision between equal masses. The secondaries emerge 90\degr\ apart\footnote{~Actually, slightly less due to a relativistic effect.} and share the incident energy, the more forward proton being the more energetic.

Quasi-elastic p-p scattering $^{16}$O(p,2p)$^{15}$N is nonelastic. Kinematics resemble scattering on free hydrogen, somewhat modified by the binding energy $E_\mathrm{B}$ required to extract the target proton and by the initial momentum of the target proton \cite{Tyren1966}. Quasi-elastic proton-neutron $^{16}$O(p,pn)$^{15}$O scattering is similar except, of course, that one secondary is neutral and travels much farther on average. (Also, the $E_\mathrm{B}$ of a neutron is greater.)

\begin{figure}[p]
\centering\includegraphics[width=2.91in,height=3.5in]{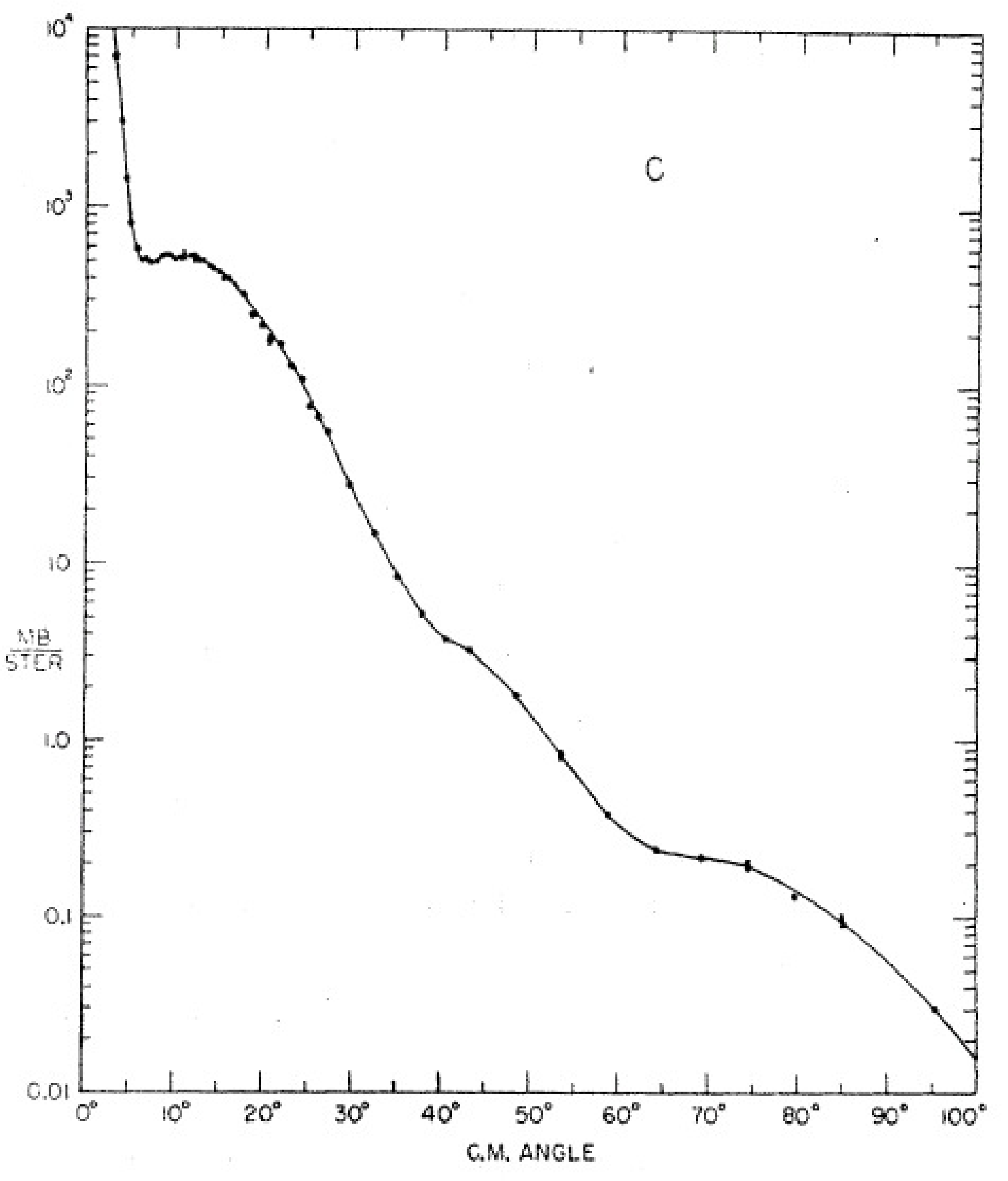} 
\caption{Gerstein et al. \cite{Gerstein1957} Fig.\,3: Elastic scattering differential cross section of 96\,MeV protons on carbon. Note (from left) the Moli\`ere single scattering tail, the Coulomb interference dip near 7\,\degr\ and scattering to large angles, with shallow diffraction minima, by the nuclear force. \label{fig:Gerstein3}}
  \end{figure}

Finally, elastic proton-nucleus $^{16}$O(p,p)$^{16}$O scattering to small angles is electromagnetic. In fact it, is identical to the Moli\`ere single scattering tail. Eventually it goes over to nuclear scattering which falls far more slowly with angle. The two regimes are separated by a {\em Coulomb interference} dip. \fig{Gerstein3} shows all this for C (O is very similar). Note that 96\,MeV is a representative KE since hard scatters can occur anywhere along the primary's path.

\subsection{Shape and Size of the Halo}
\begin{figure}[p]
\centering\includegraphics[width=4.57in,height=3.5in]{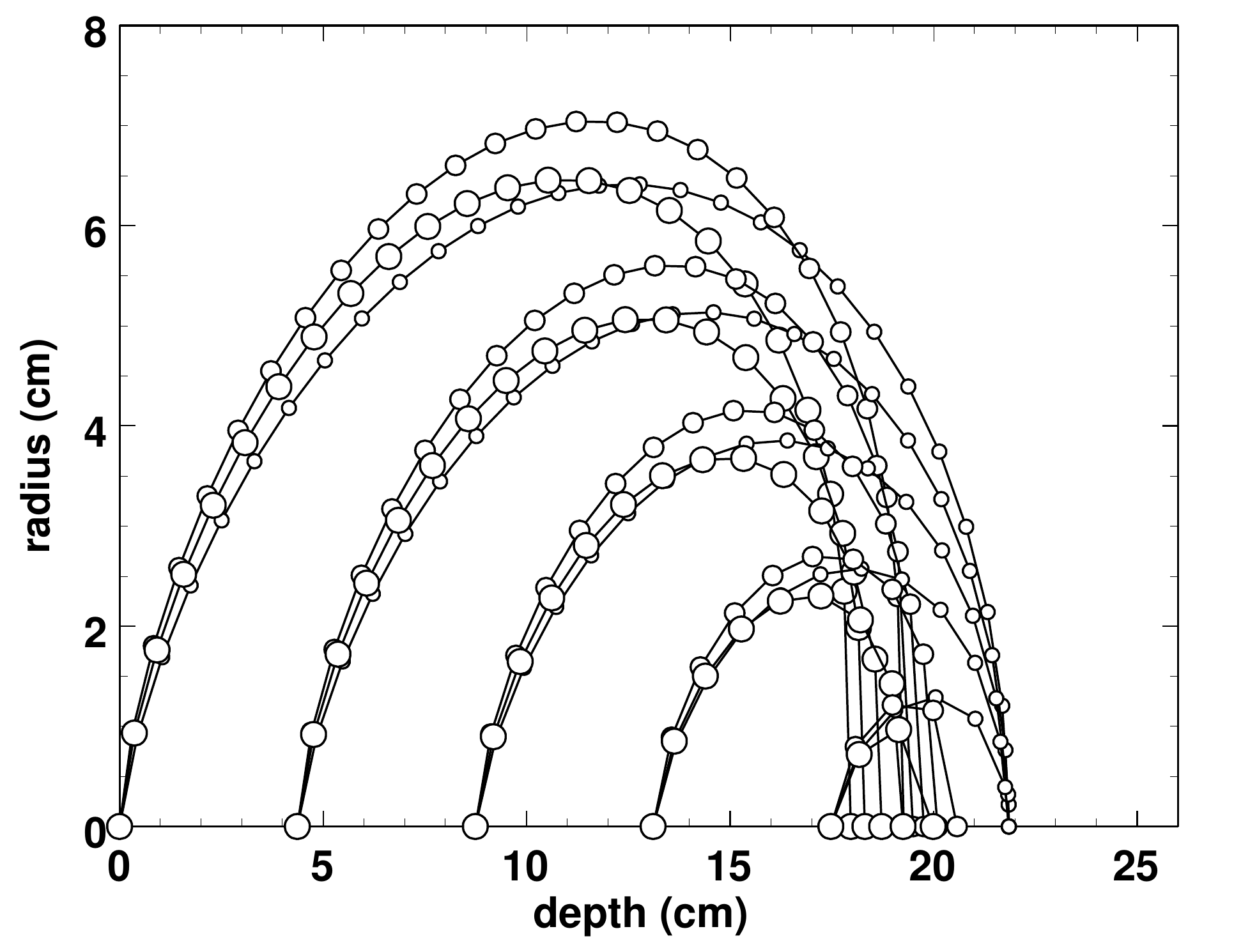}
\caption{Stopping points in water of the outgoing proton, assuming a 180\;MeV beam with reactions at five depths. Small, medium, large circles represent elastic, QE with $E_\mathrm{B}=12.4$\,MeV and QE with $E_\mathrm{B}=19.0$\,MeV. Recoil nucleus parameters are $\theta=300$\degr, $pc=75$\,MeV \cite{Tyren1966}.\label{fig:haloBump}}
  \end{figure}

Using conservation of energy and momentum one can compute the energy, and therefore the range, of the more energetic secondary in scattering from free H. The quasi-elastic case is more complicated \cite{Gottschalk2015,Gottschalk2014} but we have good values for $E_\mathrm{B}$ as well as the characteristic target proton momentum from the literature \cite{Tyren1966}. \fig{haloBump} shows, for 180\,MeV incident protons, stopping points of secondary protons from elastic scattering on free H as well as quasi-elastic scattering in O, assuming the reaction occurs at one of five depths along the primary track. The figure strongly suggests that a longitudinal scan at $\approx6$\,cm will exhibit a dose bump around mid-range.

From similar figures at other incident energies we can conclude that the radius of the halo is very roughly one-third the range of the incident beam.  

\subsection{Experiment}
Pedroni et al. \cite{pedroniPencil}, and others since then, used a PBS facility itself to investigate the halo in water, laying down hollow frames of PBs and measuring the resulting depth-dose on the central axis. By contrast, Sawakuchi et al \cite{Sawakuchi2010} used a single PB on the axis of a water tank and took radial scans with a small IC at selected depths. We adopted that more direct approach at 177\,MeV, with two important changes. First, we used a beam monitor calibrated to measure the total number of incident protons and second, we took depth scans at ten radii instead of radial scans at selected depths. The result \cite{Gottschalk2015,Gottschalk2014} was the first absolute and comprehensive measurement of the halo.

\begin{figure}[p]
\centering\includegraphics[width=4.78in,height=3.5in]{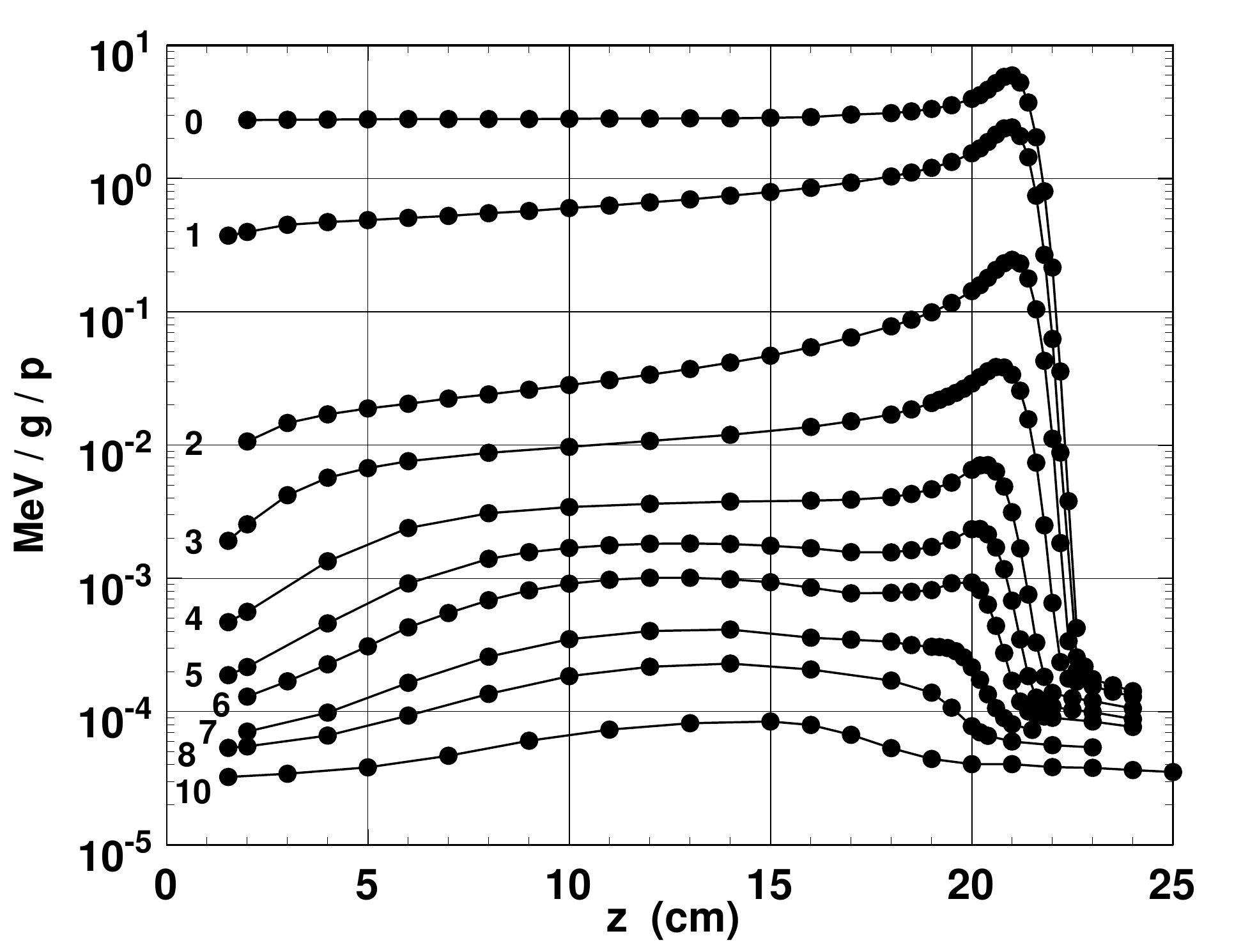}
\caption{Measured depth-dose distributions at 10 radii \cite{Gottschalk2015}. Left-hand numbers are the distance (cm) of each scan from the beam center line. The lines serve only to guide the eye. 1\,MeV/g = 0.1602\,nGy.\label{fig:dmlg}}
  \end{figure}

\begin{figure}[p]
\centering\includegraphics[width=4.66in,height=3.5in]{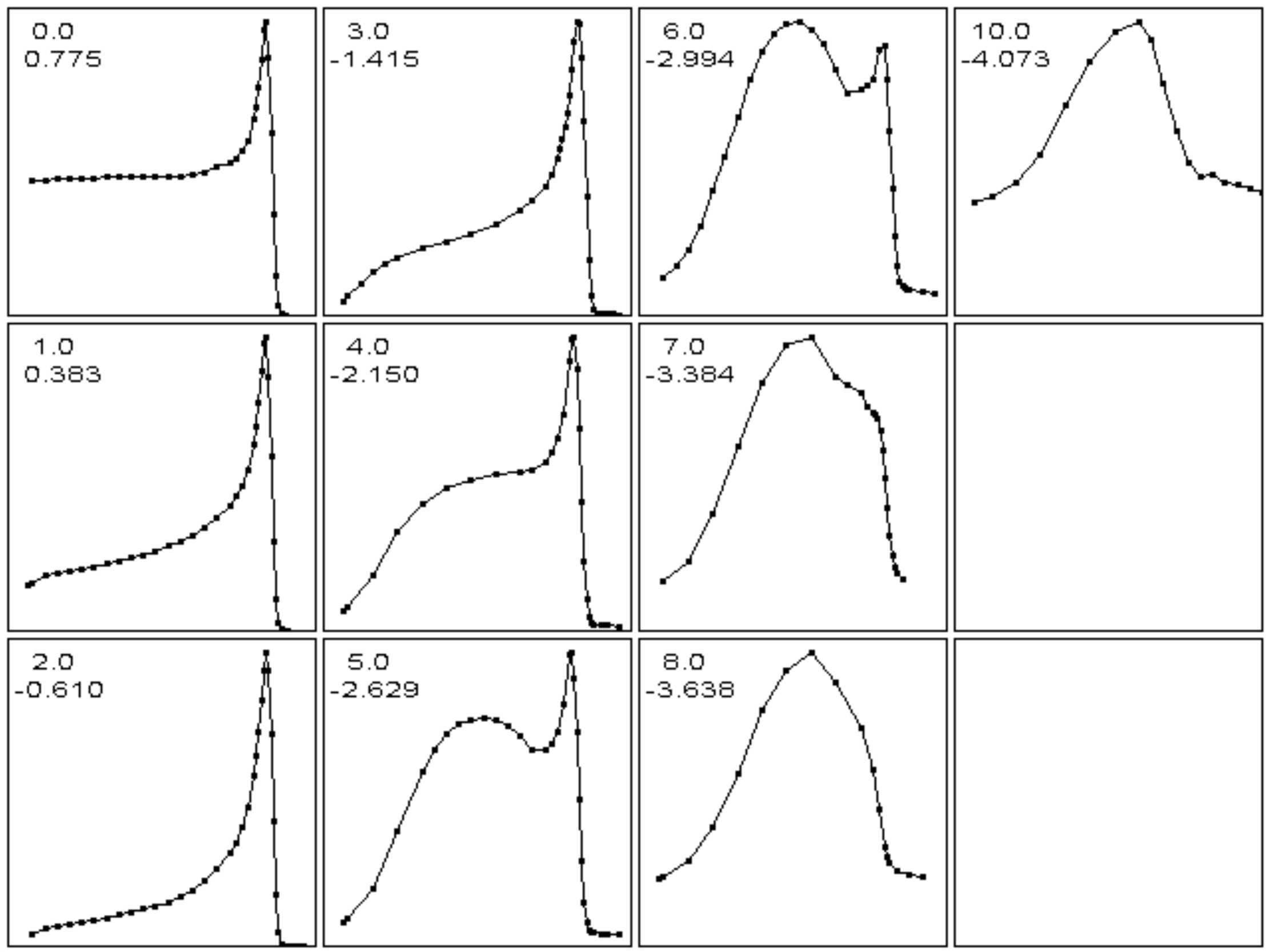}
\caption{The same data as Fig.\,\ref{fig:dmlg} in auto normalized linear form. Upper number in each frame is $r$ (cm), lower is log$_{10}$(dose/(MeV/g/p)) for the highest point in the frame.\label{fig:dmlin}}
  \end{figure}

Semilog \fig{dmlg} shows two advantages of our method. Experimentally, it requires fewer adjustments of electrometer gain and beam intensity because the dose rate in any given scan is more nearly constant. More important, it reveals features of the halo which are easily missed by selected radial scans. Linear \fig{dmlin} shows these more clearly: the midrange bump at large radii predicted by elastic and quasi-elastic kinematics, the Bragg peak at radii far larger than the Moli\`ere single scattering tail and lastly, evidence for the aura as dose well beyond the Bragg peak, and as a smooth background at large radii.

\begin{figure}[p]
\centering\includegraphics[width=4.72in,height=3.5in]{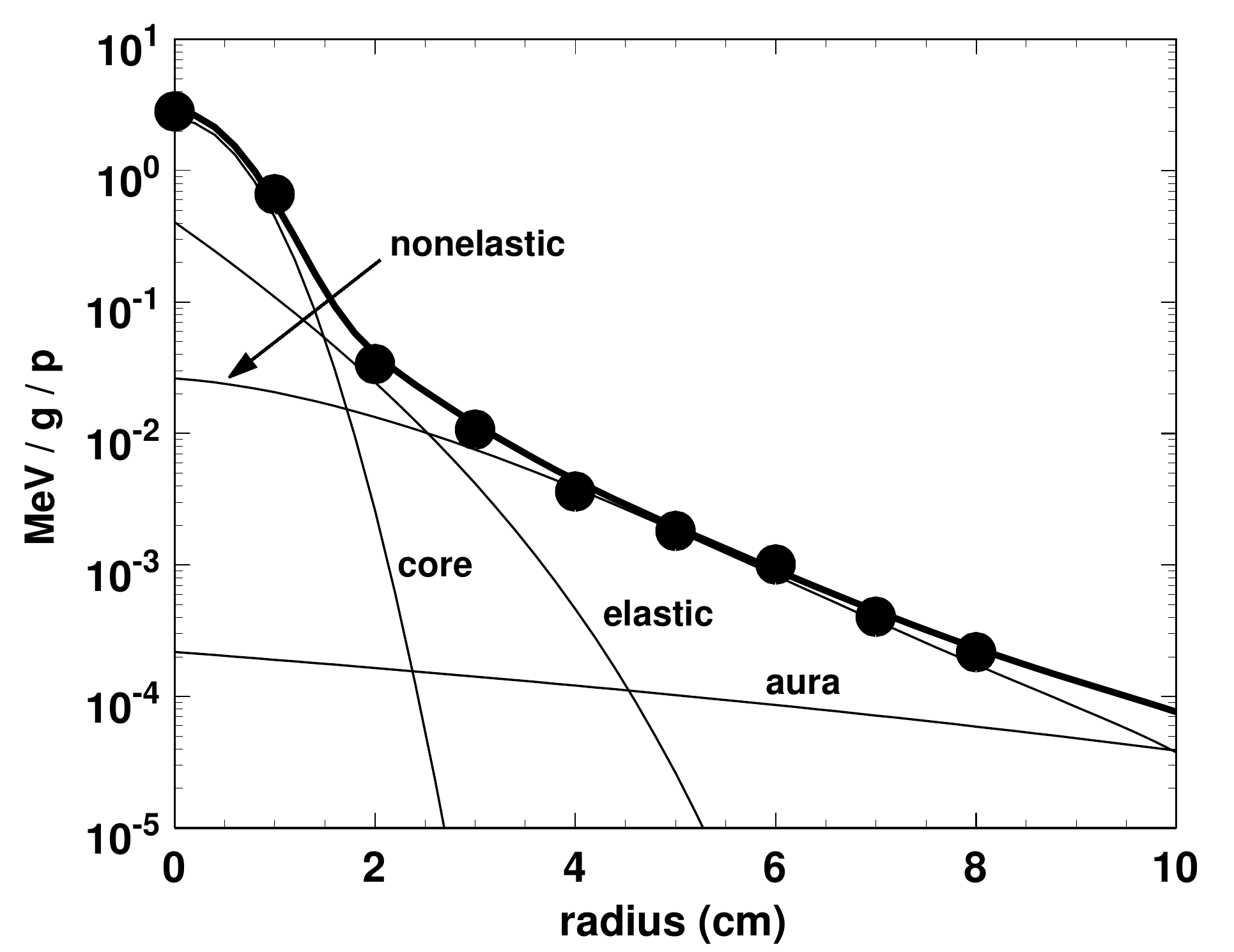}
\caption{Bold line: model dependent fit \cite{Gottschalk2015} to the transverse dose distribution at $z=12$\,cm (midrange) with experimental points (full circles). Light lines: contribution of each term in the MD fit.\label{fig:trans12}}
  \end{figure}

These data can be fit in an model-dependent fashion \cite{Gottschalk2014} which separates, at least approximately, the core, halo elastic, halo nonelastic, and aura. \fig{trans12} shows a transverse distribution at midrange. The transverse shape varies strongly with depth making it difficult to fit with any single functional form.

\subsection{Halo and Aura as Monte-Carlo Tests}
The experimental data just described provide an incisive test of Monte Carlo nuclear models because they are absolute and because the multiple EM, Moli\`ere tail, and nuclear coherent and incoherent regions, each related to a different model, are experimentally separated as well as the underlying physics permits. At this writing only Geant4 has been tested against these data. Overall, absolute agreement is remarkably good although some issues remain \cite{Hall2016}. Whether these are experimental, defects in the Geant4 model, or both, has yet to be determined.

\clearpage
\section{Bragg Curve and Effective Stopping Power}\label{sec:Bragg}
As previously noted, every dose calculation starts with
\begin{equation}D\;=\;\Phi\times \frac{S}{\rho}\end{equation}
However, we must interpret $S$ as the {\em effective} stopping power $S_\mathrm{eff}$ of a cohort of protons, not the theoretical CSDA value for a single proton, which is very large at low $T$.\footnote{~For instance, $S/\rho=115$\,MeV/(g/cm$^2$) at 3\,MeV in water (\eqn{bethe}).} The effective mass stopping power of a beam is much lower because the protons stop at slightly different depths due to range straggling, which smooths the dose out considerably.

To grasp `effective stopping power' consider this question: what is the stopping power of a 160\,MeV beam in water at the Bragg peak? Answering that by direct calculation is difficult, since we need to average over a good model of the mix of energies and stopping points at the Bragg peak. However, observe that the peak/entrance dose ratio of the Bragg curve under typical circumstances is $\approx4$. Since tabulated $-(1/\rho)\,dT/dz$ in water at 160\,MeV (the entrance energy) is $\approx5.2$\,Mev/(g/cm$^2$) the {\em effective} mass stopping power of that beam at the Bragg peak is simply $\approx4\times5.2=21$\,Mev/(g/cm$^2$).

We consider two limiting versions of $S_\mathrm{eff}$, representing extreme cases. Both are required to understand and parameterize pencil beams. The first, $S_\mathrm{em}$, is the effective stopping power of a proton beam with hard scatters turned off. As we will see, it is easy to compute but it cannot be measured because hard scatters cannot, in fact, be turned off. The second, $S_\mathrm{mixed}$, is the effective stopping power prevailing near the center of a broad beam. By contrast with $S_\mathrm{em}$ it can be measured (two ways, actually) but cannot be computed, except by Monte-Carlo, because we have no simple mathematical model for hard scatters.

\subsection{Electromagnetic Stopping Power $S_\mathrm{em}$}
$S_\mathrm{em}$ (suppressing $\rho_\mathrm{water}$) is simply the convolution of \eqn{bethe} with a Gaussian whose rms value $\sigma_\mathrm{sem}$ combines range straggling and initial beam energy spread in quadrature: 
\begin{equation}\label{eqn:convo}
S_\mathrm{em}(z)\,\equiv\,\int_{z-5\sigma}^{z+5\sigma}-\frac{dT}{dz}(z)\;G_\mathrm{1D}(z-z';\sigma_\mathrm{sem})\,dz'
\end{equation}
$G_\mathrm{1D}$ is given by \eqn{G1D} and \eqn{convo} assumes $G_\mathrm{1D}$ is negligible at $\pm5\,\sigma_\mathrm{sem}$ . 

We can perform the integral numerically using Simpson's Rule \cite{nr} but we must deal with the singularity of $dT/dz$ at end-of-range.\footnote{~Alternatively, the integral can be expressed in closed form subject to some approximations \cite{bortfeld}.} We therefore break the integral into two terms at some cutoff depth $z=z_c$\,, one term nonsingular and the other susceptible to approximation: 
\begin{equation}\label{eqn:cutoff}
S_\mathrm{em}(z)\,=\,\int_{z-5\sigma}^{z_c}-\frac{dT}{dz}(z)\;G(z-z';\sigma_\mathrm{sem})\,dz'\,+\,
  \int_{z_c}^{z+5\sigma}-\frac{dT}{dz}(z)\;G(z-z';\sigma_\mathrm{sem})\,dz'
\end{equation}
The second integrand is dominated by the singularity at $R\approx z_c$ so we approxi\-mate it by setting $z'=z_c$ and find
\begin{equation}\label{eqn:Sem}
S_\mathrm{em}(z)\,\approx\,\int_{z-5\sigma}^{z_c}-\frac{dT}{dz}(z)\;G(z-z';\sigma_\mathrm{sem})\,dz'\,+\,T(R-z_c)\;G(z-z_c;\sigma_\mathrm{sem})
\end{equation}
$T(R-z_c)$ is the residual kinetic energy at $z_c$, which we can obtain from range-energy tables. The overall result is insensitive to $z_c$.

\begin{figure}[p]
\centering\includegraphics[width=4.55in,height=3.5in]{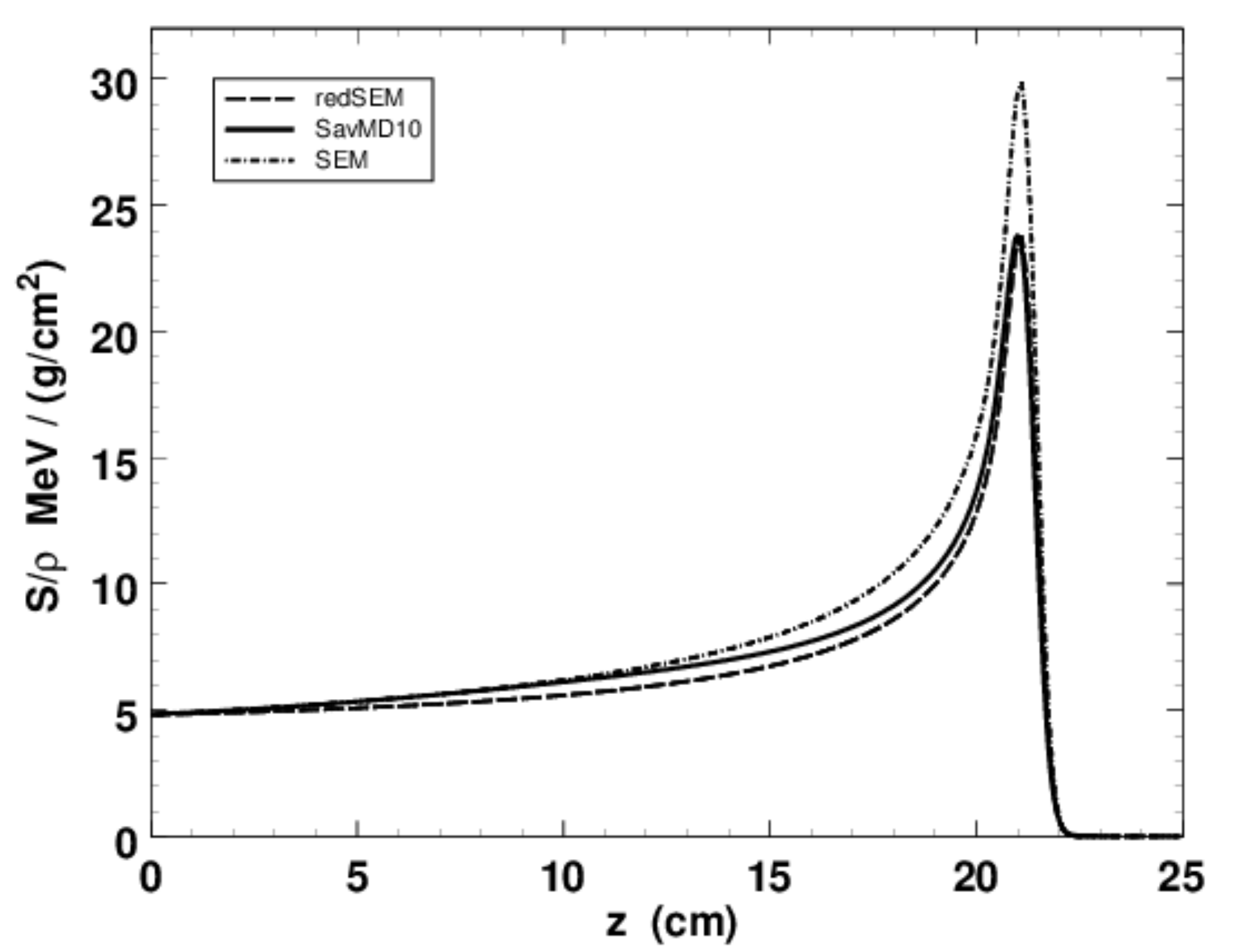}
\caption{Bragg curves in water. Dot-dash: EM mass stopping power $S_\mathrm{em}/\rho$. Solid: mixed mass stopping power. Dashed: $S_\mathrm{em}/\rho$ times a factor $(1-c\,z)$ with $c$ adjusted to equalize the peak values.\label{fig:S_em}}
  \end{figure}

$S_\mathrm{em}$ is plotted in \fig{S_em} (dot-dash line). The peak is greater than we observe in a real proton beam because the CSDA ignores hard scatters (nuclear interactions), assuming energy loss is entirely due to multiple soft EM interactions with atomic electrons. When hard scatters are turned on, fewer primaries reach the peak, a corresponding amount of energy is either moved upstream (charged secondaries) or entirely outside the region of interest (neutral secondaries), and the solid line results. We discuss this next.

\subsection{Transverse Equilibrium and $S_\mathrm{mixed}$}
\begin{figure}[p]
\centering\includegraphics[width=4.095in,height=3.5in]{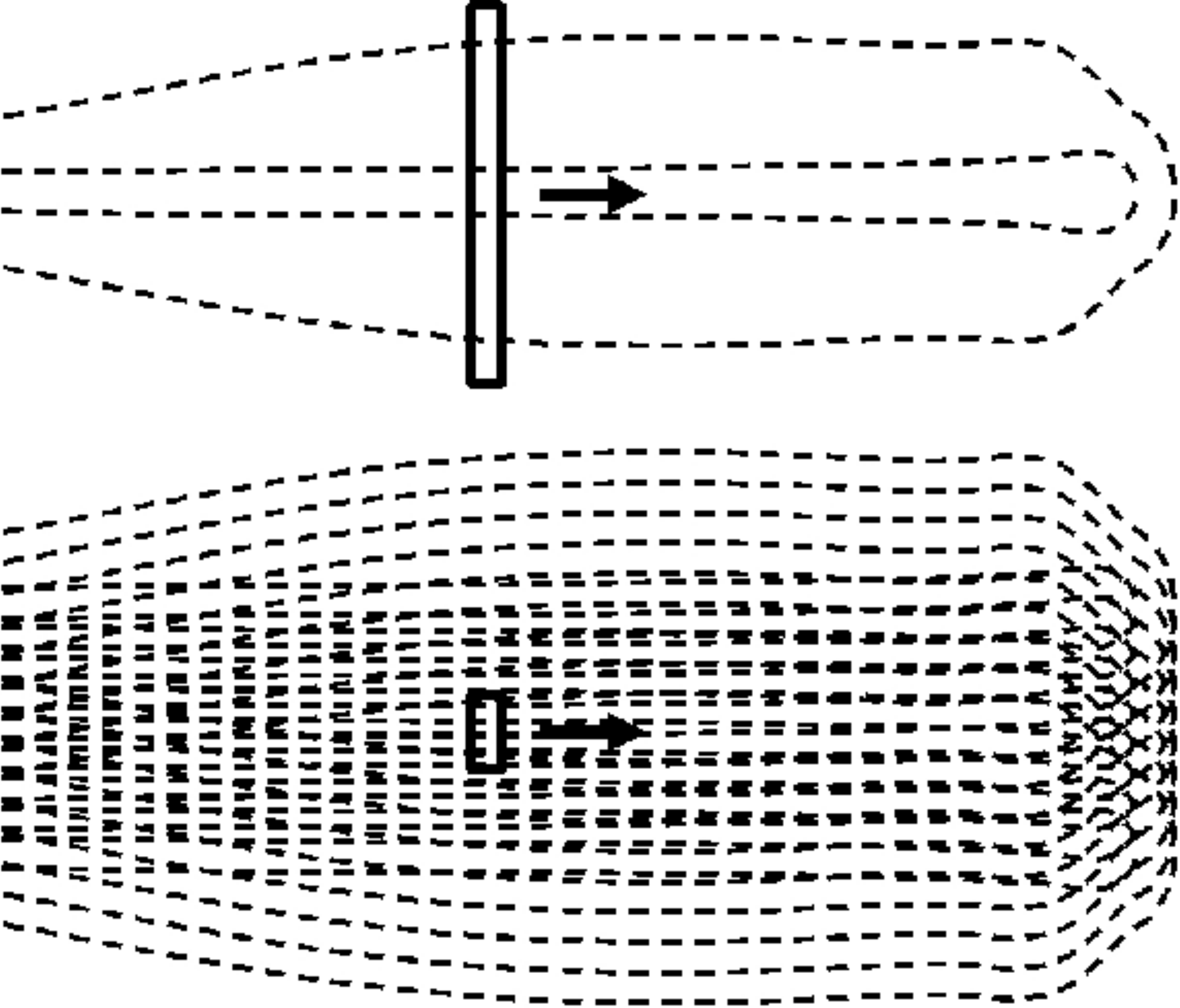}
\caption{Top: depth scan of a single pencil beam with a Bragg peak chamber (BPC). Bottom: depth scan of a broad beam with a small IC.\label{fig:equilibrium}}
  \end{figure}

Suppose we apply a large number of PBs close together and do a depth scan with a small IC (\fig{equilibrium} bottom). Because of both multiple EM and hard scatters, not all the protons initially aimed at the IC actually reach it. However, those that scatter away from the IC are exactly compensated by those from neighboring PBs that scatter into it. Therefore, near the center of a sufficiently broad beam in a uniform medium, the exact transverse position of the IC does not matter. This condition is called {\em transverse equilibrium}. As we have seen, the characteristic radius $R_\mathrm{halo}$ of hard scatters is much greater than that of multiple EM scatters, and therefore governs the beam size required for transverse equilibrium.

If transverse equilibrium holds, the mix of particles at a given depth, both primaries and secondaries, is independent of the transverse coordinates. To calculate dose under those conditions, $S_\mathrm{mixed}$ is appropriate. As \fig{equilibrium} shows, it can be measured either by using a small IC in a sufficiently broad beam, or a sufficiently large `pancake' IC (a so-called `Bragg peak chamber', BPC) straddling a single PB. We now show that mathematically.

\subsection{Measuring $S_\mathrm{mixed}$}\label{sec:MeasuringSmixed}

We wish to show that the integrated dose $D$ to water per incident proton, measured by a BPC straddling a single pencil beam, equals the local dose to water, divided by the fluence in air, on the axis of a sufficiently broad  uniform beam. 

The average energy $dE_1$ per proton deposited in a disk of thickness $dz_1$ and radius $R_\mathrm{halo}$ by many protons entering a large water tank near the axis (\fig{equilibrium} top) is
\begin{equation}\label{eqn:dE1}
dE_1(z)\;=\;\left(\int_0^{2\pi}\int_0^{R_\mathrm{halo}}D_1(r,z)\;r\;dr\;d\phi\right) \rho\;dz_1
\end{equation}
Now let the tank be exposed, instead, to a broad uniform parallel proton beam of fluence in air $\Phi$. The energy $dE_n(z)$ deposited in a small disk of thickness $dz_1$ and radius $R_\mathrm{dosim}$ by $n$ protons at the tank entrance directed at the disk (\fig{equilibrium} bottom) is
\begin{equation}\label{eqn:dEn}
dE_n(z)\;=\;\left(\int_0^{2\pi}\int_0^{R_\mathrm{dosim}}D_n(r,z)\;r\;dr\;d\phi\right) \rho\;dz_1
\;=\;\pi\;R_\mathrm{dosim}^2\;D_n(0,z)\;\rho\;dz_1
\end{equation}
In the second step we have assumed transverse equilibrium (EM and nuclear) near the axis, so that $D_n$ is independent of $r$ there. Because the mix of stopping powers is the same in both discs, and they have the same thickness, it follows that
\begin{equation}\label{eqn:dE1n}
dE_1(z)\;=\;dE_n(z)/n
\end{equation}
Transverse equilibrium also plays a role in \eqn{dE1n}. Protons initially directed at the small dosimeter do not deposit all their energy in it. However, the shortfall is exactly compensated by energy from protons {\em not} directed at the dosimeter. We thus find
\begin{equation}
\int_0^{2\pi}\int_0^{R_\mathrm{halo}}D_1(r,z)\;r\;dr\;d\phi\;=\;D_n(0,z)/\bigl(n/(\pi\,R_\mathrm{dosim}^2)\bigr)
  \;=\;D_n(0,z)\,/\,\Phi
\end{equation}
Dose per fluence is a mass stopping power (\eqn{D}), which we have called $S_\mathrm{mixed}/\rho_\mathrm{water}$.

In practice, either method has its problems. Commercial BPCs are not quite large enough at the higher energies \cite{Gottschalk2015}. Scanning a BPC in a water tank can be unwieldy, so it is sometimes incorporated into a water column \cite{ptw} which must of course be big enough. If the BPC is too small, Monte-Carlo derived corrections can be applied. 

The small IC/broad beam method with a single dosimeter suffers from the fact that {\em all} the PBs must be laid down to measure a single point on the depth-dose. That may be prohibitively slow. Multilayer ICs (MLICs) have been developed to address this \cite{MLICBG} and there is at least one commercial version, the IBA `Zebra' \cite{iba}.

\subsection{Parameterizing the Nuclear Halo}
The PSI group \cite{pedroniPencil} first pointed out the need to take the energy in the halo into account.\footnote{~The contribution of the aura to the high dose region is negligible.} Their parameterization of the halo deserves special attention because it set the pattern for all subsequent PBS treatment planning systems (TPSs) and is, in our opinion, flawed. It calls for unnecessary measurements which, unfortunately, have become enshrined in practice.

We merely outline the argument here. See \cite{Gottschalk2015,Gottschalk2014} for details and references. PSI's $T(w)$ is identical to our $S_\mathrm{mixed}(z)/\rho_\mathrm{water}$. Their parameterization (their Eq.\,(7)), in our notation and assuming cylindrical symmetry, reads 
  \begin{equation}\label{eqn:PSI}
D(r,z)\;=\;\left[(1-f_\mathrm{NI}(z))\times G_\mathrm{2D}^\mathrm{P}(r;\sigma_\mathrm{P}(z))+f_\mathrm{NI}(z)\times G_\mathrm{2D}^\mathrm{NI}(r;\sigma_\mathrm{NI}(z))\right]\times\frac{S_\mathrm{mixed}}{\rho_\mathrm{water}}
  \end{equation} 
the familiar (dose) = (fluence)$\times$(mass stopping power). The fluence, in square brackets, is divided into a primary or core term (P) and a secondary or `nuclear interaction' term (NI). $f_\mathrm{NI}$, an adjustable positive function of $z$ and incident energy, governs the attrition of primaries. It reduces the core and increases the halo by the same amount. $\sigma_\mathrm{NI}(z)$, another adjustable function of $z$ and energy, is the rms width of a 2D Gaussian describing the fluence in the halo, whereas $\sigma_\mathrm{P}(z)$ is the rms width of a 2D Gaussian describing the core,  not adjustable but, instead, found by transport (Fermi-Eyges) theory.

We claim that associating $S_\mathrm{mixed}(z)$ with the core is incorrect; it should be $S_\mathrm{em}(z)$. Primaries suffer only multiple soft EM scatters, by definition, and not the hard scatters included in $S_\mathrm{mixed}(z)$. Dose from hard scatters ends up in the halo, not the core! A red flag is the fact that $S_\mathrm{mixed}(z)$ has the primary attrition function built-in, so the explicit $(1-f_\mathrm{NI}(z))$ is duplicative, whereas $S_\mathrm{em}$ does not have built-in primary attrition. What mass stopping power to associate with the {\em halo} is a more complicated question, for which see \cite{Gottschalk2014}. 

\begin{figure}[p]
\centering\includegraphics[width=4.70in,height=3.5in]{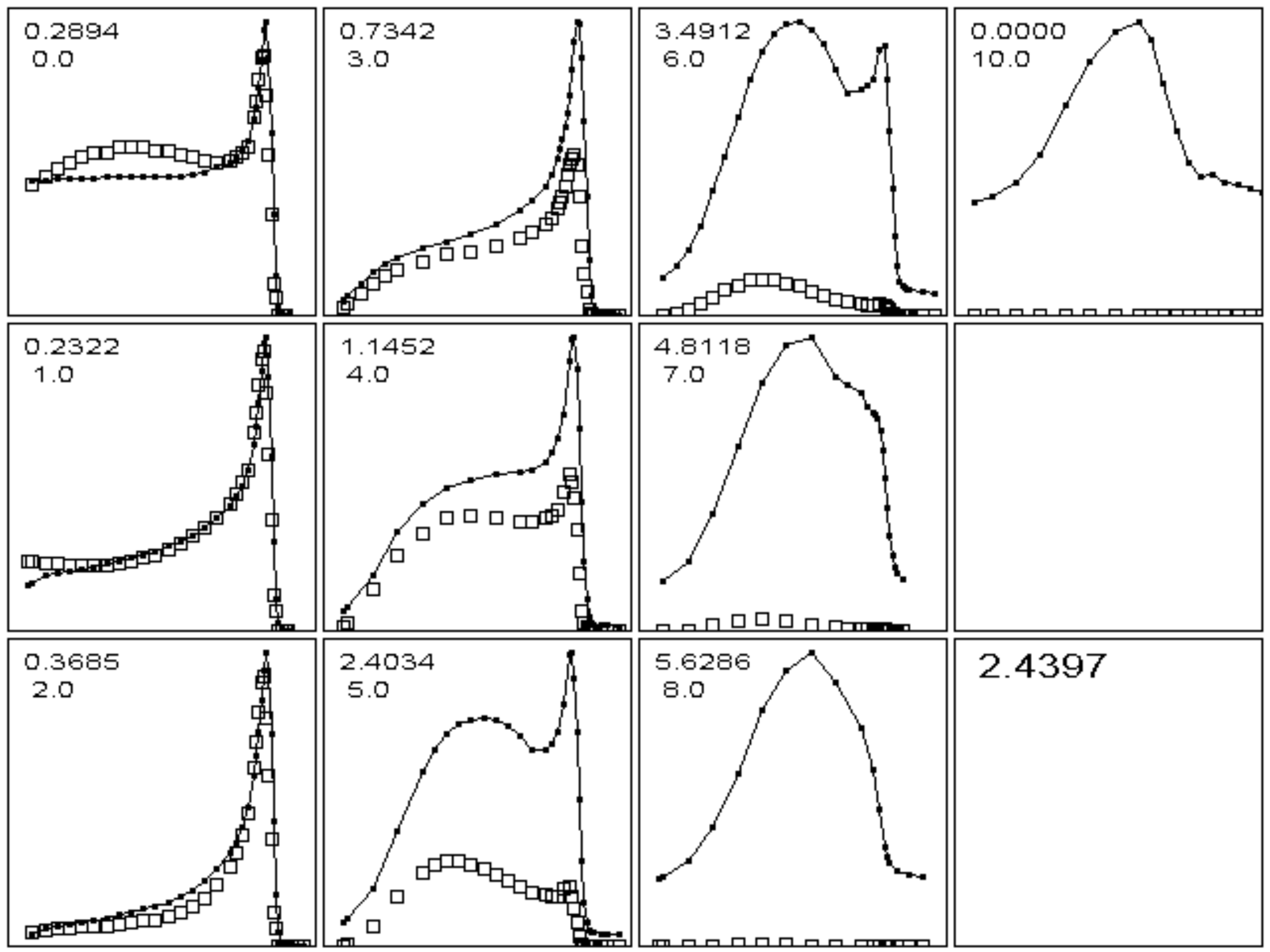}
\caption{Our data with the PSI fit (absolute comparison) adjusted to our initial beam size.\label{fig:PedLinFitEM}}
  \end{figure}

\fig{PedLinFitEM} compares our data \cite{Gottschalk2015} with the PSI fit \cite{pedroniPencil} at 177\,MeV, with $\sigma_\mathrm{P}$ adjusted to our beam size. Agreement in the core is acceptable, except for the excess at midrange ($r=0$) from using $S_\mathrm{mixed}$ instead of $S_\mathrm{em}$ cf. \fig{S_em}. A more obvious but less consequential problem is the Gaussian describing secondaries, which falls far too rapidly (\fig{PedLinFitEM}). PSI themselves characterized this as a `first preliminary estimate' and others (e.g. \cite{Bellinzona2015}) have used radial functions with longer tails. However, we caution that the transverse dose depends strongly on depth \cite{Gottschalk2015}, and therefore, no single shape can be regarded as typical.

Our objection to \eqn{PSI} is not that it causes errors in treatment planning, since any commercial TPS has enough adjustable parameters to compensate for the erroneous use of $S_\mathrm{mixed}$. Rather, it is the misconception that it is necessary to measure $S_\mathrm{mixed}$, leading in turn to commercial and unnecessary BPCs and MLICs. To compound the problem, the commercial BPCs are not quite large enough at the higher energies. That has lead to experimental \cite{Anand2012} and Monte-Carlo \cite{Clasie2012} papers to correct for lost signal in an unnecessary measurement.

In summary, the nuclear halo at a given energy is best mapped by measuring the depth-dose of a single PB at selected radii with a small dosimeter, preferably with an absolute beam monitor. The radial integral of dose,  $T(w)$ or IDD or $S_\mathrm{mixed}/\rho_\mathrm{water}$, is inessential. The most efficient parameterization of the core and halo for treatment planning purposes is, in our view, still an open question.

\subsection{Energy Dependence of Bragg Curves}
\begin{figure}[p]
\centering\includegraphics[width=4.49in,height=3.5in]{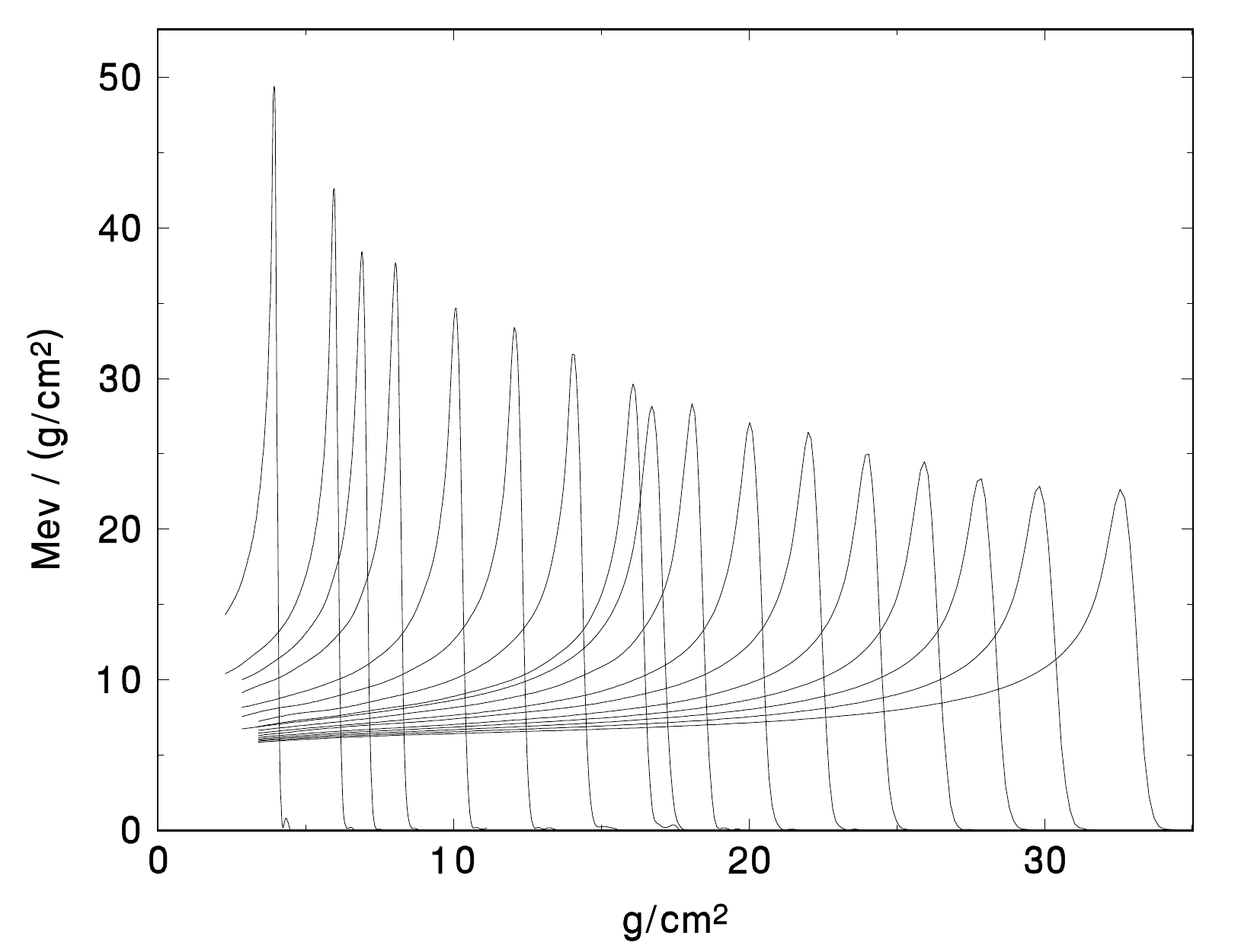}
\caption{Measured Bragg curves from 69 to 231\,MeV (data courtesy D. Prieels, Ion Beam Applications s.a.).\label{fig:bpscans1}}
  \end{figure}

Since straggling dominates the Bragg peak, and is a constant fraction of range, it follows that BPs taken at lower energies are sharper (\fig{bpscans1}). When we spread the dose in depth by adding BPs of different ranges (range modulation) we can either use degraders\footnote{~A {\em degrader} is any slab whose primary purpose is to reduce energy.} or change the machine energy. In the former case, the preceding discussion (\sect{rangeStraggling}) means that component BPs are just pulled back versions of the deepest one. In the latter case, the sharpening of the component BPs must be taken into account.

\subsection{The Disappearing Bragg Peak}\label{sec:disappearing}
\begin{figure}[p]
\centering\includegraphics[width=4.45in,height=3.5in]{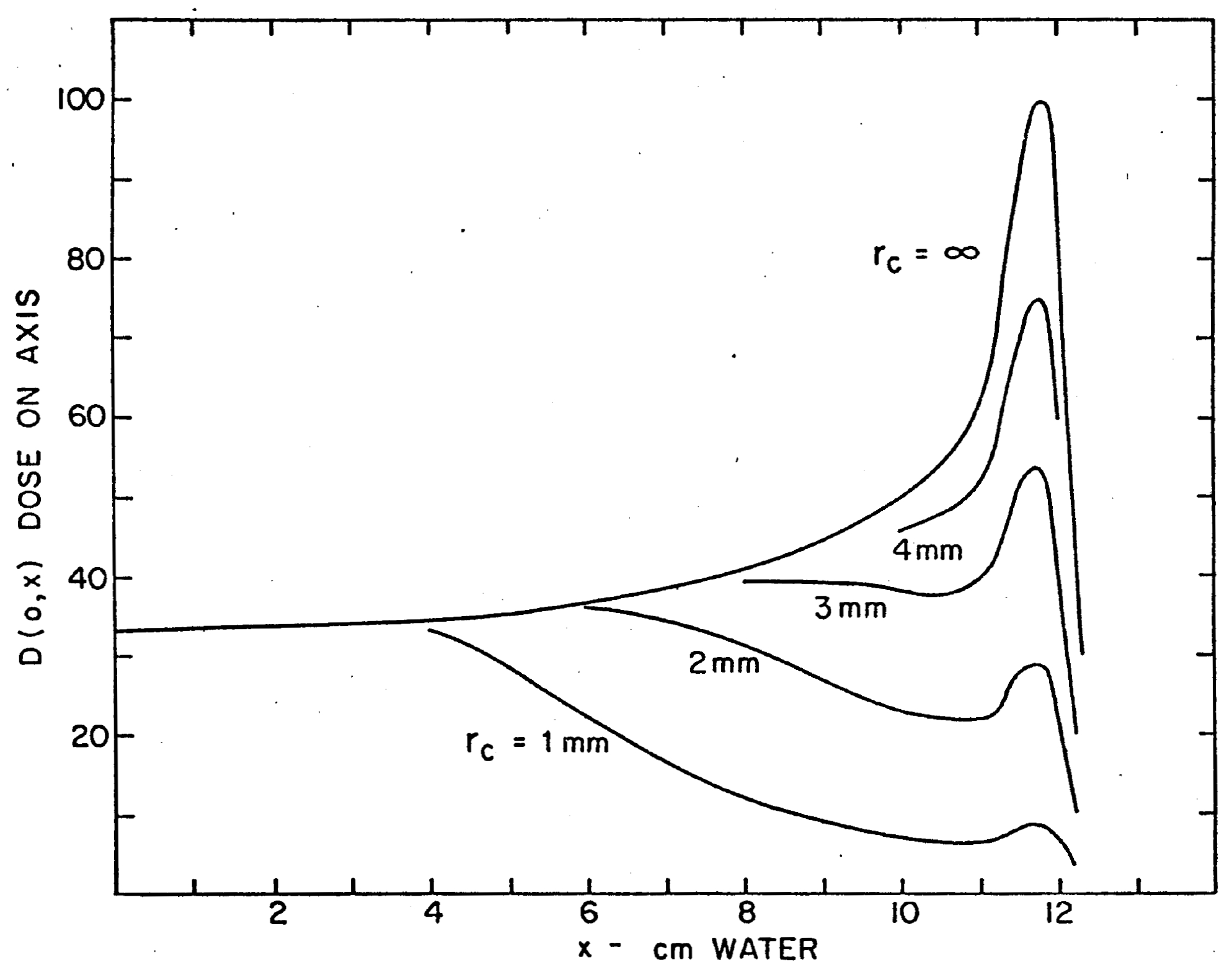}
\caption{Fig.\,7 of Preston and Koehler \cite{preston}: relative dose on the axis of a uniform circular proton beam of initial range 12~cm water and radius $r_c$ at the collimator. The curve for $r_c=\infty$ is experimental, the others are calculated.\label{fig:pk7}}
  \end{figure}

Rather than our two limiting cases, large dosimeter/small beam and small dosimeter/large beam, what happens with a small dosimeter along the axis of a small beam? Again, we use $D=\Phi\times S/\rho$. As depth increases, energy decreases and stopping power increases. At the same time, fluence decreases because MCS spreads the beam out (the 2D Gaussian grows wider) and there is no compensating in-scattering from neighboring PBs. The decrease in fluence ultimately wins out, and the Bragg peak disappears. First predicted by Preston and Koehler \cite{preston} (\fig{pk7}), this has been confirmed experimentally many times. It is a limiting factor in treating small deep fields.

\subsection{Nuclear Buildup}
If the depth-dose in a broad beam entering a water tank from air or vacuum is measured with a small dosimeter, a nuclear buildup of a few percent in the first cm or so is observed. This was first reported by Carlsson and Carlsson \cite{carlsson}. They noted that the buildup was smaller than expected.

It should be clear from the foregoing discussion that observed buildup depends on both beam size and detector size. With a small beam and a small detector, for instance, there is no buildup (\fig{dmlin}). Hard scatters occur, but their dose appears at larger radii. Full buildup will be observed only with a small detector in a beam sufficiently large for nuclear equilibrium. Probably the beam (size not given) in \cite{carlsson} was too small.

\clearpage
\section{Looking Ahead}
As promised, we have discussed the three basic interactions of radiotherapy protons with matter: EM stopping, EM multiple scattering, and hard scatters. That still leaves us a long way from computing the dose in realistic situations. For that, we need a {\em transport theory}, namely a general way of computing fluence. Let us take a quick look ahead. 

\subsection{Fermi-Eyges Theory}\label{sec:FE}
Enrico Fermi, in unpublished lecture notes, computed the transverse displacement, the angle, and their correlation, for cosmic rays traversing the earth's atmosphere \cite{rossi}. He used the Gaussian approximation to MCS. Leonard Eyges \cite{eyges} added the effect of energy loss, which Fermi had ignored. Eyges assumed an ideal beam entering a single homogeneous slab, and built into his theory a scattering power ($T_\mathrm{FR}$  in \cite{scatPower2010}) which, it turns out, is the worst possible choice. Later, the theory found application in electron radiotherapy\footnote{~Ironically, protons satisfy the approximations of FE theory far better than do electrons.} and was generalized to non-ideal incident beams, multiple homogeneous slabs, and arbitrary scattering powers. For a detailed account please see \cite{transport2012}.

In Fermi-Eyges (FE) theory, the terrain is discretized by `$z$-planes' perpendicular to the nominal beam direction. These may be actual slab boundaries, or may simply be introduced for computational purposes. At each such plane, a cylindrically symmetric Gaussian pencil beam (PB) is fully characterized by ten parameters: one for the total proton charge the PB carries, three for the position $x,y,z$ and two for the direction $\theta_x,\theta_y$ of the PB's central axis, one for the nominal proton energy or an equivalent such as residual range in water or $pv$, and three `Fermi-Eyges moments' related to the PB's transverse size, angular divergence, and emittance.\footnote{~The emittance, a measure of the correlation between position and angle, is the most subtle. If position and angle are perfectly correlated the emittance is zero, even if the beam has finite width and angular divergence. If they are completely uncorrelated, the emittance is maximal.}

To `transport' a PB through a step $z\rightarrow z'$ is to find these parameters at $z'$ given their values at $z$. Fermi-Eyges theory is simply a prescription for doing that. It follows that we can, in the Gaussian approximation, find the parameters of any pencil beam at any depth in mixed-slab geometry by a relatively simple computation.

That is a big step forward. However, the resulting fluence distribution (and therefore the dose) is Gaussian, and only rarely does that correspond to a useful problem. (Think of the complicated dose distribution inside a patient.) Most problems involve {\em transverse} as well as longitudinal heterogeneities. Before tackling that, we take a short detour. 

\subsection{The Preston and Koehler Manuscript}\label{sec:PK}
William M. Preston (1910-1989) and Andreas M. (`Andy') Koehler (1930-2015) were among the pioneers of proton radiotherapy. Coming from a high energy rather than a radiothera\-py background, they were unaware of FE theory. However, they used an analogous method in a manuscript \cite{preston} which to this day is the best theoretical and experimental study of the evolution of an ideal PB in a homogeneous slab. Unfortunately it was rejected for publication. Proton radiotherapy was deemed uninteresting at the time (around 1968). 

Because it solves the same problem as FE theory, their theory must in some way be equivalent. In \cite{transport2012} we rederive their most important results in FE language, and extend them to heavy ions. Here we merely state them.

First: {\em the rms transverse spread $\sigma_x(R)$ at end-of-range is proportional to the range $R$}. The constant of proportionality is
\begin{equation}\label{eqn:sigxoR1}
\frac{\sigma_x(R)}{R}\;=\;\frac{E_s\,z}{2\,(pv)_\mathrm{R/2}}\sqrt{\frac{R}{X_S}}
\end{equation}
which despite appearances is very nearly independent of $R$. $E_s=15.0$\,MeV, $z$ is the particle charge number, $pv$ is evaluated at the $T$ value corresponding to $R/2$ (cf. \eqn{pv}), and $X_S$ is the scattering length  of the material \cite{scatPower2010}. In Lexan, for instance, $\sigma_x(R)=0.021\, R$. Values of $\sigma_x(R)/R$ and $\rho X_S$ for many other materials can be found in \cite{transport2012}.

\begin{figure}[p]
\centering\includegraphics[width=3.64in,height=3.5in]{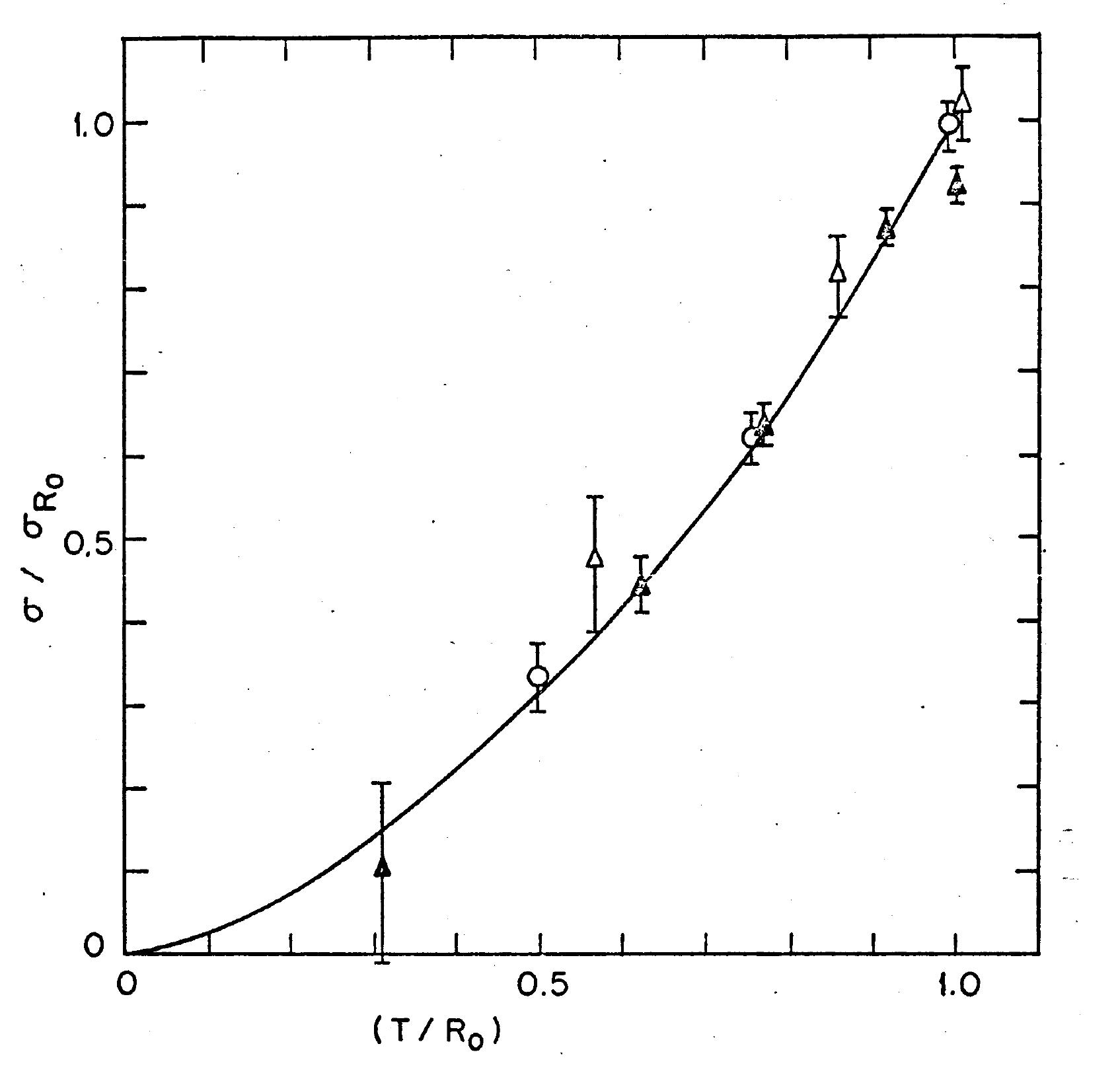} 
\caption{Facsimile of Preston and Koehler Fig.\,17, captioned ``Dimensionless plot of the standard deviation due to scattering versus depth of penetration $\ldots$ Open triangles are experimental results for 112\,MeV protons on aluminum; solid triangles for 158\,MeV protons on aluminum; open circles for 127\,MeV protons on water."\label{fig:pk17}}
  \end{figure}

Second: at any lesser depth $z<R$,
\begin{equation}\label{eqn:PKuniversal}
\frac{\sigma_x(z)}{\sigma_x(R)}\;=\;\left[2\,(1-t)^2\ln\left(\frac{1}{1-t}\right)
  +\,3\,t^2-2\,t\right]^{1/2}\hbox{\quad,\quad}t\;\equiv\; z/R
\end{equation}
which, with Eq.\,\ref{eqn:sigxoR1}, {\em completely describes beam spreading in a homogeneous slab for any heavy charged particle beam at any energy in any material}. Eqs.\,\ref{eqn:sigxoR1} and \ref{eqn:PKuniversal} assume an ideal incident beam, so measurements of $\sigma_x(z)$ must be corrected for initial beam size, divergence and emittance. These are often significant in beams designed for PBS. \fig{pk17} is Preston and Koehler's experimental confirmation of \eqn{PKuniversal} for two materials and two energies. 

Third: the {\em disappearing Bragg peak}, covered previously (\sect{disappearing} and \fig{pk7}).

\subsection{A Comprehensive Pencil Beam Algorithm}\label{sec:PBA}
The following description of a novel pencil beam algorithm (PBA), using FE theory, is a very brief outline. For a full description please see \cite{Gottschalk2016}. 

The archetypal forward problem of proton radiotherapy physics is computing the dose everywhere in a known heterogeneous terrain irradiated by known proton beams. For routine treatment planning, that computation needs to be fast and therefore deterministic. Like other deterministic algorithms, PBA relies on the fact that the high-dose region is well approximated even if we ignore hard scatters (nuclear interactions), that is, if we consider just the primary protons. Hard scatters can be put in later as a correction.

Monte-Carlo methods transport individual protons using models of basic interactions. Deterministic methods group protons into bundles or `pencil beams', transported using FE theory. As stated in \sect{FE}, any PB can be transported correctly through any number of {\em longitudinal} heterogeneities. It is {\em transverse} heterogeneities that cause difficulties.

A brass collimator furnishes an extreme example. Depending on their coordinates, protons making up a PB may find themselves either in brass or in air. However, FE theory can only transport the {\em entire} PB, and only through one material or the other. If the PB central axis is just inside brass, FE will choose brass (`central axis' or CAX approximation) and {\em all} the protons represented by the PB will stop. Otherwise, FE will choose air and {\em all} the protons will pass through, scattered and slowed only by a few cm of air. If the incident PB represents a great many protons, a serious error results either way.

We therefore break up the mother PB into smaller daughter PBs, separated transversely. That resolves the brass/air ambiguity for some of the daughters, and the others represent fewer protons so the error is less serious. The breakup is repeated as often as necessary.

For rapid convergence, PBA combines two ways of breaking up PBs. In {\em redefinition} \cite{shiu} the entire cohort of PBs is replaced at one or more selected $z$-planes. In {\em dynamic splitting} \cite{kanematsuSplit}, individual PBs are split, and only when necessary.

Besides combining redefinition with recursive dynamic splitting, PBA has other novel features. 

First and foremost, it is a calculation {\em from first principles}. That is, it starts directly at the level of established laws of physics and does not use empirical model and fitting parameters. All it requires is an accurate physical description of the incident beams, the beam line (if any), and the patient. Thus, in a passive beam spreading problem, it does not start with the beam's effective source position and size. Instead, these follow, as they should, from the physical parameters of the beam line.

Second, PBA makes no distinction between passive beam spreading and PBS. In passive spreading, the beam line is more complicated and there is only one incident beam. In PBS the beam line is simpler or absent, and there are many incident beams. 

Third, all objects (beam line, collimators, patient) are represented the same way, as stacks of homogeneous or heterogeneous slabs. There are no magical `beam defining' planes. Collimator scatter (protons interacting with the collimator without stopping in it) emerges in a natural way. 

Last, every material is associated with its correct stopping and scattering powers, rather than being treated as a variant of water. The concept of `radiological path length' is not used. That allows PBA to address problems involving e.g. titanium implants.

\begin{figure}[p]
\centering\includegraphics[width=4.67in,height=3.5in]{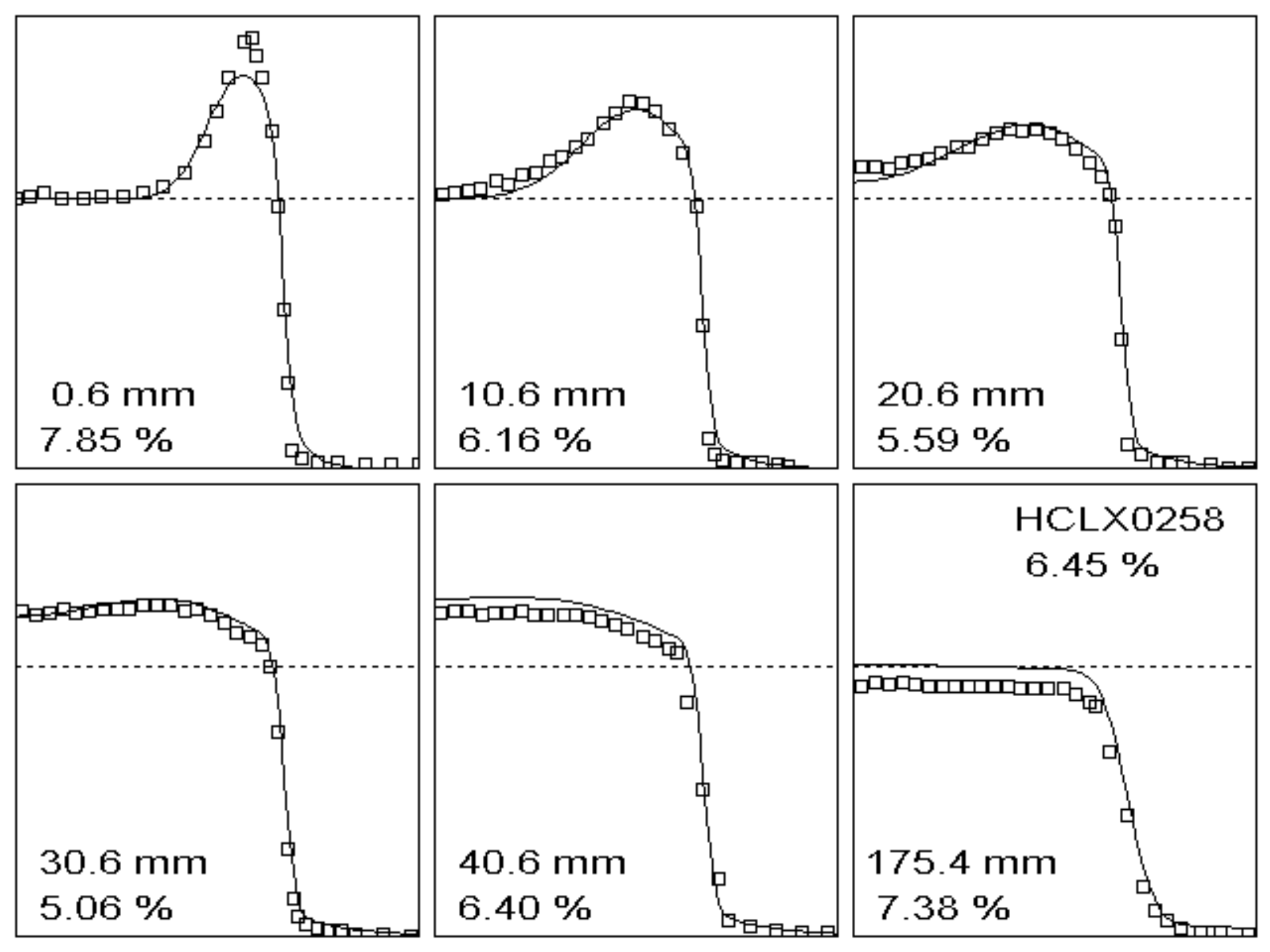}
\caption{Transverse dose in air at six distances (top number) from the downstream face of a brass collimator with a 9.88\,mm radius hole. Percentages give the rms deviation of the PBA calculation (line) from experiment (squares). 1665\,K PBs were processed and 531\,K left the collimator.\label{fig:HCLX0258}}
  \end{figure}

As an example, \fig{HCLX0258} is a PBA computation of collimator scatter (not handled at all by other PB algorithms). It takes 1.4\,minutes on a single laptop computer (Lenovo T400 running Intel Fortran under Windows\,7). The same problem using TOPAS/Geant4  takes many hours on a multi-core computer (see \cite{Gottschalk2016}).

\section{Acknowledgements}
We thank Harvard University and the Physics Department for generous and sustained support, and Harald Paganetti for valuable comments on the first draft of this writeup.

\appendix

\clearpage\section{Acronyms}\label{sec:Acronyms}
\setlength{\tabcolsep}{10pt}
\begin{center}
\begin{tabular}{ll}
\multicolumn{1}{c}{acronym}&           
\multicolumn{1}{c}{description}\\           
\noalign{\vspace{8pt}}
BP&Bragg peak\\
BPC&Bragg peak (ionization) chamber\\
CSDA&continuous slowing down approximation\\
EM&electromagnetic\\
FC&Faraday cup\\
FE&generalized Fermi-Eyges (theory)\\
IC&ionization chamber\\
IDD&integral depth dose (PSI's $T(w)$)\\ 
KE&kinetic energy\\
MCS&multiple Coulomb scattering\\
MGH&Massachusetts General Hospital (Boston, Massachusetts, USA)\\
MLFC&multilayer Faraday cup\\
MLIC&multilayer ionization chamber\\
MP&measuring plane\\
MS&manuscript\\
PBA&the author's pencil beam algorithm\\
PBS&pencil beam scanning\\
PMFC&poor man's Faraday cup\\
POI&point of interest\\
PSI&Paul Scherrer Institut (Z\"urich, Switzerland)\\
TPS&treatment planning system\\
\end{tabular}
\end{center}

\clearpage\section{Symbols}\label{sec:Symbols}
\setlength{\tabcolsep}{10pt}
\begin{center}
\begin{tabular}{lll}
\multicolumn{1}{c}{symbol}&           
\multicolumn{1}{c}{description}&           
\multicolumn{1}{c}{units}\\           
\noalign{\vspace{8pt}}
$A$&area&cm$^2$\\
$A_0,A_1,A_2$&Fermi-Eyges moments&rad$^2$, cm\;rad, cm$^2$\\
$c$&speed of light&299\,792\,458\,m/s = 0.984\,ft/ns\\
$D$&dose&Gy $\equiv$ J/Kg\\
&& 1\,MeV/g = 0.1602\,nGy\\
$E$&total energy&MeV\\
$e$&quantum of charge&1.602$\times10^{-10}$\,nC\\
$i$&electric current&nA\\
$mc^2$&proton rest energy&938.272\,MeV\\
$N$&number of protons\\
$N_\mathrm{A}$&Avogadro constant&$6.022141\times10^{23}$\,mol$^{-1}$\\
$pc$&proton momentum&MeV\\
$pv$&proton momentum times speed&MeV\\
$q$&charge&nC\\
$R$&range&cm\\
$\rho R$&mass range&g/cm$^2$\\
$S$&stopping power $\equiv-dT/dz$&MeV/cm\\
$S/\rho$&mass stopping power $\equiv-dT/(\rho dz)$&MeV/(g/cm$^2$)\\
$t$&time&s\\
&mass target thickness in MCS derivation&g/cm$^2$\\
$T$&kinetic energy&MeV\\
$T_\mathrm{dM}$&differential Moli\`ere scattering power&rad$^2$/cm\\
$v$&speed&cm/s\\
$X_0$&radiation length&cm\\
$X_\mathrm{S}$&scattering length&cm\\
$x,y,z$&transverse coordinates&cm\\
\noalign{\vspace{10pt}}
$\alpha$&fine structure constant&1/137.035\\
&Born parameter in Moli\`ere theory\\
$\beta$&v/c\\
$\theta_0$&Gaussian width parameter&rad\\
$\theta_x,\theta_y$&projected angles&rad\\
$\Phi$&fluence$\;\equiv dN/dA$&protons/cm$^2$\\
$\Phi_q$&charge fluence$\;\equiv dq/dA$&nC/cm$^2$\\
$\rho$&density&g/cm$^3$\\
\end{tabular}
\end{center}

\clearpage\section{Gaussians}\label{sec:Gaussians}
\subsection{1D}
When protons pass through a thin scatterer they suffer deflections in angle $\theta_x,\theta_y$ (multiple Coulomb scattering or MCS) that are independent and very nearly Gaussian. Assume an incident beam of $N$ protons is ideal (no transverse or angular spread) and initially directed along the $z$ axis ($x=y=0$). Assume also a screen or `measuring plane' (MP) a distance $L$ downstream (\fig{MCSdemo}). To a very good approximation the probability per proton of finding $\theta_x$ in $d\theta_x$ irrespective of $\theta_y$ is 
\begin{equation}
G_\mathrm{1D}(\theta_x;\theta_0)\,d\theta_x\;=\;\frac{1}{\sqrt{2\pi}\;   \theta_0}\;e^{\textstyle{-\frac{1}{2}\left(\frac{\theta_x}{\theta_0}\right)^2}}\;d\theta_x
\end{equation}
$\theta_0$ is the {\em width parameter} of the Gaussian.

It is easier and ultimately more useful to focus our attention on the MP. Each angular deflection $\theta_x$ becomes a transverse deflection $x=L\,\theta_x\;$\footnote{~Proton MCS angles are always small in the sense $\tan(\theta)\approx\sin(\theta)\approx\theta$. We have ignored energy loss and MCS in the air.} The probability of observing a proton on the MP at $x$ in $dx$ irrespective of $y$ is
\begin{equation}\label{eqn:G1D}
G_\mathrm{1D}(x;x_0)\,dx\;=\;\frac{1}{\sqrt{2\pi}\;x_0}\;
   e^{\textstyle{-\frac{1}{2}\left(\frac{x}{x_0}\right)^2}}\;dx\end{equation}
It is normalized: the probability that the proton is somewhere in the MP is
\begin{equation}\int_{-\infty}^\infty\;G_\mathrm{1D}(x;x_0)\;dx\;=\;1\label{eqn:G1norm}\end{equation}
and its {\rm rms} width, using \eqn{G1norm}, is
\begin{equation}\sigma_x\;\equiv\;<x^2>^{1/2}
  \;=\;\left(\int_{-\infty}^\infty\;x^2\;G_\mathrm{1D}(x;x_0)\;dx\right)^{1/2}\;=\;x_0\end{equation}
This result is often absorbed into the notation, that is, \eqn{G1D} is written with $\sigma_x$ in place of $x_0$. It gives the probability per proton of a count in a sufficiently long vertical strip detector of width $dx$, located at the MP.

\subsection{2D (Cylindrical)}
Suppose instead we have a small 2D detector of area $dx\,dy$. In amorphous materials, angular deflections in $x$ and $y$ are independent and equally probable. The probability per proton of finding $x$ in $dx$ and $y$ in $dy$ is therefore
\begin{equation}
  G_\mathrm{1D}(x;x_0)\;dx\times G_\mathrm{1D}(y;y_0)\;dy\;=\;
  \frac{1}{2\pi\;x_0\;y_0}\;e^{\textstyle{-\frac{1}{2}\left[\left(\frac{x}{x_0}\right)^2+
  \left(\frac{y}{y_0}\right)^2\right]}}\,dx\,dy
\end{equation}
or, in cylindrical coordinates $r^2\equiv x^2+y^2$, $r_0\equiv x_0=y_0$, $dx\,dy=r\;dr\;d\phi$
\begin{equation}\label{eqn:G2D}
  G_\mathrm{2D}(r,\phi;r_0)\;r\;dr\;d\phi\;=\;
  \frac{1}{2\pi\;r_0^2}\;e^{\textstyle{-\frac{1}{2}\left(\frac{r}{r_0}\right)^2}}\;r\;dr\;d\phi
\end{equation}
This is still normalized, the probability that the proton is somewhere in the MP being
\begin{equation}
  \int_0^{2\pi}\int_0^\infty\;G_\mathrm{2D}(r,\phi;r_0)\;r\;dr\;d\phi\;=\;1
\end{equation}
However, the {\em rms} value of $r$ is
\begin{equation}
  \sigma_r\;\equiv\;<r^2>^{1/2}
  \;=\;\left(\int_0^{2\pi}\int_0^\infty\;r^2\;G_\mathrm{2D}(r,\phi;r_0)\;r\;dr\;d\phi\right)^{1/2}\;=\;\sqrt{2}\;r_0
\end{equation}
leading many authors to write \eqn{G2D} as 
\begin{equation*}
  G_\mathrm{2D}(r,\phi;\sigma_r)\;r\;dr\;d\phi\;=\;
  \frac{1}{\pi\;\sigma_r^2}\;e^{\textstyle{-\left(\frac{r}{\sigma_r}\right)^2}}\;r\;dr\;d\phi
\end{equation*}
which can get confusing. In reading any work on MCS you need to decide which convention for the 2D Gaussian width parameter the authors are using. Otherwise, you'll be off by a factor $\sqrt{2}=1.414$. Our convention is $r_0$ or \eqn{G2D}, in which the 1D and 2D width parameters have the same numerical value.

Finally, we cite an exact property of cylindrical Gaussians given by Preston and Koehler \cite{preston}. The efficiency (fraction of protons inside $r_{xx}$) is
\begin{equation}\label{eqn:eps}
\epsilon\;\equiv\;\int_0^{2\pi}\int_0^{r_{xx}}\;G_\mathrm{2D}(r,\phi;r_0)\;r\;dr\;d\phi\;=\;
  1\;-\;e^{-\textstyle{\frac{1}{2}\left(\frac{r_{xx}}{r_0}\right)^2}}\;=\;
  1\;-\;G_\mathrm{2D}(r_{xx})/G_\mathrm{2D}(0)
\end{equation}
For instance, if we use the Gaussian out to where it is 95\% of its central value, down 5\%, the efficiency is also 5\%. Low efficiency leads to low dose rate, limiting the usefulness of single scattered beams to small fields. Eventually, that lead to the development of double scattering to improve the dose rate for large fields. Nowadays magnetic scanning, with even higher efficiency, is supplanting double scattering for medium and large fields.

\clearpage\section{Relativistic Single-Particle Kinematics}\label{sec:Kinematics}
At 160\,MeV $v/c\approx0.5$ so radiotherapy protons are relativistic. Although nonrelativistic equations are often adequate, the exact equations given here are preferable.

A proton beam is usually characterized by its kinetic energy $T$, but we occasionally need its speed $\beta c$ or momentum $pc$ or the quantity $pv$ which governs MCS. The next three formulas can be found in any introductory book on special relativity:\,\footnote{~$E$ is {\em total} energy. Medical physicists often use $E$ for {\em kinetic} energy but then $E=mc^2$, the most famous equation of the twentieth century, makes no sense. There is also precedent in the medical physics literature for $T$ e.g. \cite{icru49}. Equations in this section apply to any particle of rest energy $mc^2$.}
\begin{equation}\beta\;\equiv\;\frac{v}{c}\;=\;\frac{pc}{E}\label{eqn:beta}\end{equation}
\begin{equation}E\;\equiv\;T\;+\;mc^2\label{eqn:E}\end{equation}
\begin{equation}E^2\;=\;(pc)^2\;+\;(mc^2)^2\label{eqn:Esq}\end{equation}
 Define $\tau$ (reduced KE) and  $\xi$ (reduced $pv$) by
\begin{equation}\tau\equiv T/(mc^2)\label{eqn:tau}\end{equation}
\begin{equation}\xi\equiv pv/(mc^2)\label{eqn:xi}\end{equation}
From the first three equations we find
\begin{equation}\beta^2\;=\;\frac{2+\tau}{(1+\tau)^2}\;\tau\label{eqn:betasq}\end{equation}
\begin{equation}pv\;=\;\frac{2+\tau}{1+\tau}\;T\label{eqn:pv}\end{equation}
\begin{equation}(pc)^2\;=\;(2+\tau)\;mc^2\;T\label{eqn:pcsq}\end{equation}
Over the radiotherapy range $3\le T\le300$\,MeV the coefficient of $T$ in \eqn{pv} varies from 2 to 1.76 so $pv$ can be thought of as roughly $2T$. 
We occasionally require the inverse of \eqn{pv} which is 
\begin{equation}T=0.5\;mc^2\,\left(\sqrt{(2-\xi)^2+4\xi}-(2-\xi)\right)\end{equation}

\clearpage\section{Typical Problems in Proton Radiotherapy Design}\label{sec:TypicalProblems}
This section discusses how the theories of stopping and MCS discussed earlier can be used to solve some of the simpler problems arising in beam line design for both passive spreading and PBS.

\subsection{Conventions and Notation}
Assume the beam travels from left to right nominally in the $z$ direction. $y$ is up and $x,y,z$ form a right-handed frame. In a beam line consisting of multiple elements (homogeneous or heterogeneous slabs), numeric subscripts refer to the upstream or entrance face. Thus $z$ is the coordinate of an arbitrary POI along the beam direction, $z_1$ is the $z$ coordinate of the upstream face of the first element, $T_3$ is the proton kinetic energy entering the third element and, if there are only three elements, $T_4$ is the KE leaving the beam line. 

It is convenient to write down solutions in a psuedo-code resembling Fortran.
Assume that we have a set of range-energy tables  and a handy interpolation routine at our disposal, as well as a routine that computes MCS in any of its variants.\footnote{~These are among the routines or `tasks' offered by LOOKUP \cite{lookup}.} Specifically, 
\begin{itemize}
\item\texttt{Densty(m1)} returns density $\rho_1$ (g/cm$^3$)
\item\texttt{Range(t1,m1)} returns mass range $(\rho R)_1$ (g/cm$^2$) given incident KE $t_1$ (MeV)
\item\texttt{Energy(r1,m1)} returns incident energy $t_1$ (MeV) given mass range \texttt{r1} (g/cm$^2$) 
\item\texttt{Dedx(t1,m1)} returns mass stopping power $(S/\rho)_1$ (MeV/(g/cm$^2$)) given $t_1$
\item \texttt{Theta0(v,g1,t2,m1)} returns $\theta_0$ (radian) given mass target thickness \texttt{g1} (g/cm$^2$) and outgoing kinetic energy \texttt{t2} (MeV). (We specify {\em outgoing} KE so that the problem is always well posed. If $t_2\le0$, $g_1$ is assumed to be stopping thickness. Remember that even then, half the protons emerge!)
\end{itemize}
Throughout, \texttt{m} stands for a material code e.g. \texttt{`LEXAN'} and \texttt{v} for a variant code e.g. \texttt{`HIGHLAND'} or \texttt{`MOLIERE'}. 

\subsection{Stopping}
A simple forward problem: 160\,MeV protons pass through 5 g/cm$^2$ Lexan. Find their outgoing energy. Most obvious but incorrect:
\begin{verbatim}
t2 = t1 - 5. * Dedx(160.,`LEXAN') = 135.19 MeV
\end{verbatim}
The correct solution is
\begin{verbatim}
t2 = Energy(Range(160.,`LEXAN') - 5.) = 133.67 MeV
\end{verbatim}
The first procedure is wrong, and always gives too high an answer, because stopping power increases with depth. It is nearly correct for sufficiently thin slabs, but consistently using the second method avoids having to test for that.

A simple inverse problem: find the thickness of lead required to reduce 160\,MeV protons to 150\,MeV. Solution: 
\begin{verbatim}
thickness = (Range(160.,`LEAD') - Range(150.,`LEAD')) / Densty(`LEAD')
          = 0.331 cm
\end{verbatim}
In short, manipulate \texttt{Range(T,M)} and \texttt{Energy(R,M)} and avoid using \texttt{Dedx}, in order to get the right answer for stopping problems regardless of the slab thicknesses involved.

\subsection{Water Equivalence}
The {\em water equivalent} of a slab arbitrary material is the thickness of water that gives the same energy loss. It can be found using
\begin{verbatim}
rrL  = Range(t1,lead) - Densty(lead)*thkL
t2   = Energy(rrL,lead)
rrW  = Range(t2,water) 
thkW = (Range(t1,water) - rrW)/Densty(water)	
\end{verbatim}
The `beam line' is a single slab of either lead (L) or water (W) and \texttt{rr} stands for `residual range'. Obviously L and W can be any other two materials. For instance, if we have a MLFC consisting of copper plates separated by thin Kapton sheets we might want to know the copper equivalent of Kapton so as to express the MLFC measured range in terms of equivalent copper thickness.

Water equivalence should be used with caution. Only if a material is water-like (e.g. plastic) is its water equivalent independent of energy. Example: the second scatterer in a common beam spreading system \cite{iba}, 0.62\,cm Pb, is equivalent to 3.556\,cm water at 200\,MeV and  3.355\,cm at 70\,MeV, a difference of 2\,mm.

Furthermore, even though the foregoing routine is correct for any slab thickness (unlike some other published methods \cite{zhang,waterEquivBG}) it is only as good as range-energy tables, say 1$-$2\%. If treatment depth depends on it, water equivalent must be {\em measured}, not computed. Simply measure the shift in the Bragg peak when the material under test is placed in front of a water tank.

All this has consequences for design. In the MGH neurosurgery beam, the Pb scatterers are placed upstream of the Lexan degraders. Thus they always see nearly the same energy, and therefore have a well-defined water equivalent, which need only be measured once for each degrader. If the Pb were downstream of the Lexan, its water equivalent would depend somewhat on the amount of Lexan inserted.

\subsection{Multiple Coulomb Scattering}\label{sec:MCSprobs}
Find the outgoing MCS angle \texttt{a2} ($=\theta_0$) from a degrader of mass thickness \texttt{g1} g/cm$^2$, or (inverse problem) the \texttt{g1} required to produce a desired \texttt{a2}. The first is easy with our standard routines:
\begin{verbatim}
t2 = Energy((Range(t1,m1) - g1),m1)
a2 = SQRT(a1**2 + Theta0(v,g1,t2,m1)**2)
\end{verbatim}
Notice that we have added any initial MCS angle \texttt{a1} (such as beam divergence) quadratically to that introduced by the degrader. Though it may be incorrect in principle \cite{lynch}, adding MCS angles in quadrature works well enough in practice.

The inverse problem is harder because we cannot solve Highland's equation analytically for target thickness. However, any inverse problem can be solved numerically by trial and error if the solution of the forward problem is known, and with modern computers that is remarkably fast. Formally, we seek a solution to
\begin{equation}\label{eqn:Fofx}
F(x)-C\;=\;0
\end{equation}
where $F$ is a procedure which yields $\theta_0$ given $x=$\;\texttt{g1} (and other parameters such as \texttt{t1} and \texttt{m1}) and $C$ is the desired value of $\theta_0$. \eqn{Fofx} can be solved by one of a number of standard procedures called {\em root finders} \cite{nr}, all of which are basically trial and error using different strategies to determine the next trial. Any root finder requires us to specify the range $a\le x\le b$ in which a root is to be sought or, in computer language, to `bracket' the root. Sometimes finding appropriate brackets can be the hardest part of the job, but \texttt{a = 0} and \texttt{b = Range(t1,m1)} will work here. It goes without saying that we have omitted a great many programing details: how to pass other parameters such as \texttt{t1} to \texttt{F}? how to respond to errors? and so forth. 

Using the technique just sketched, LOOKUP can find (for instance) the mass thickness of Pb which will yield $\theta_0=10$\,mrad for 130\,MeV protons . The answer is $0.265$\,g/cm$^2$, and as a bonus LOOKUP tells us the outgoing energy is $129.21$\,MeV. Check the answer against Highland's formula \eqn{Highland}.

\subsection{Binary Degraders}
\begin{figure}[p]
\centering\includegraphics[width=4.51in,height=3.5in]{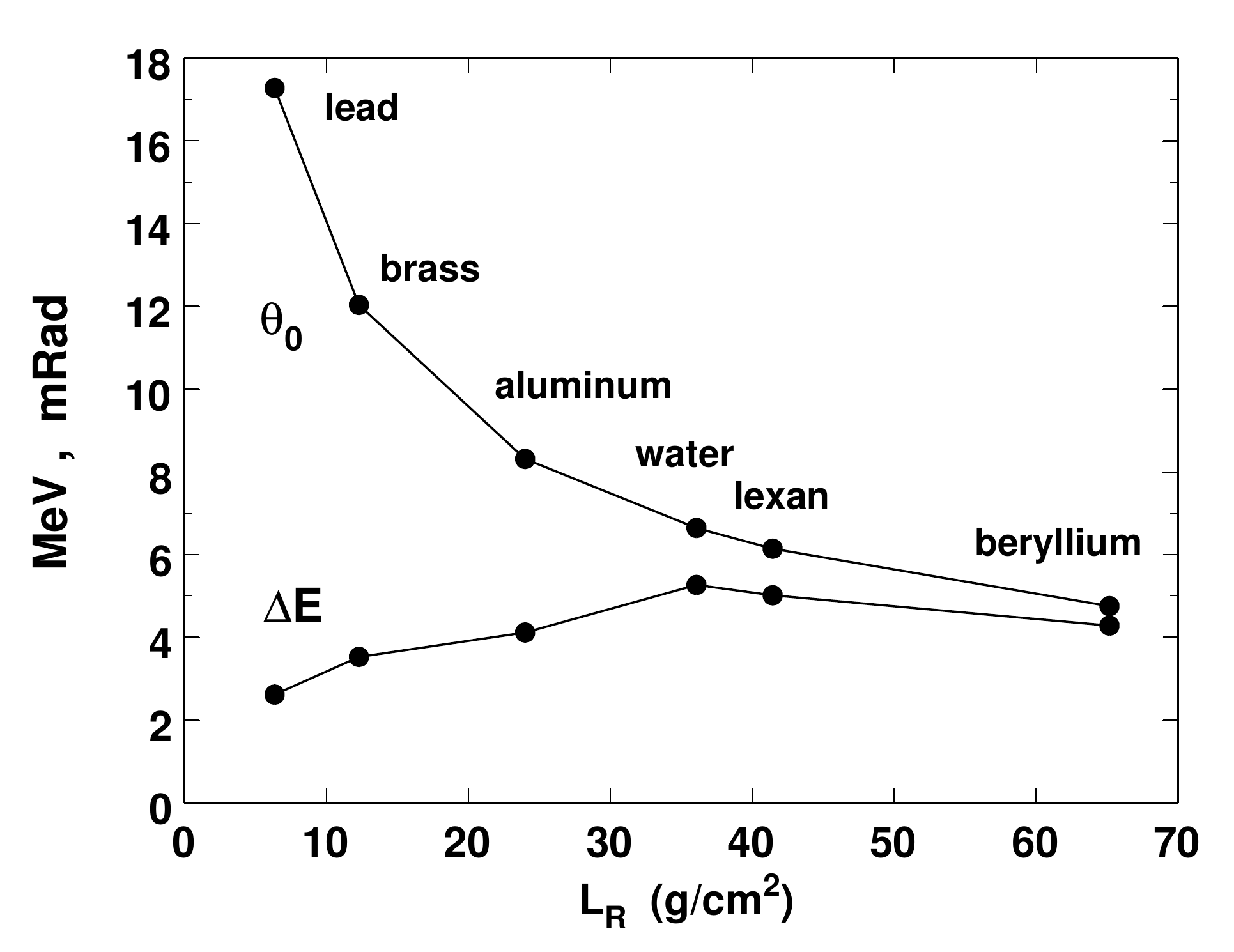}
\caption{Multiple scattering angle and energy loss for 160\,MeV protons traversing 1\,g/cm$^2$ of various materials \cite{icru49}.\label{fig:hiZloZ}}
  \end{figure}

Starting with an ideal beam at some energy, we have learned how to reduce the energy by a desired amount or how to produce a desired MCS angle. It may happen, often in passive beam spreading but occasionally also in PBS, that we wish to modify {\em both} the energy and the MCS angle. We use the fact that stopping and scattering vary in opposite ways as we traverse the periodic table (\fig{hiZloZ}). If we want scattering with minimum energy loss lead (Pb) is best, whereas for maximum energy loss and minimum scattering beryllium (Be) is best (but many plastics are acceptable).

\begin{figure}[p]
\centering\includegraphics[width=4.58in,height=3.5in]{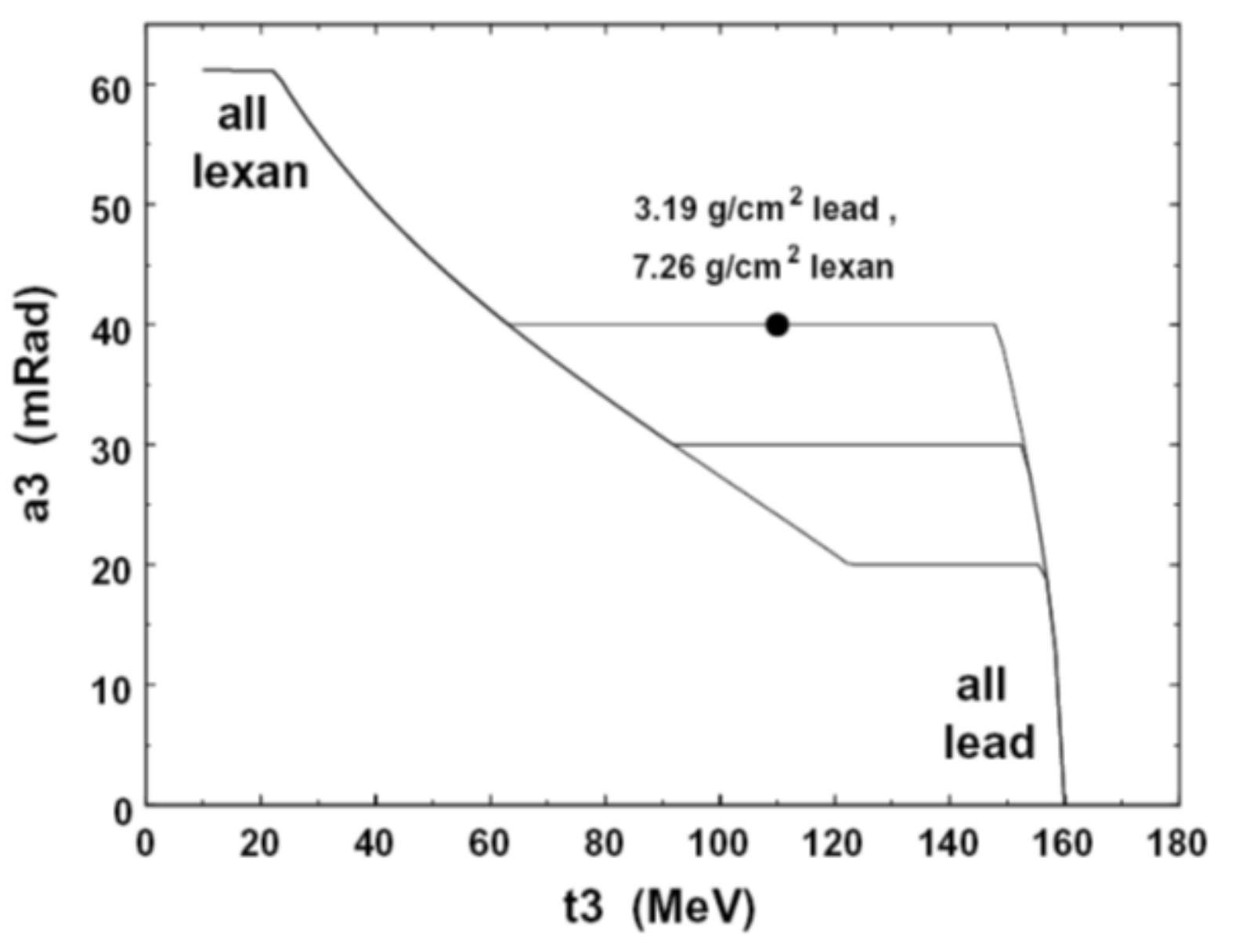}
\caption{Energy loss and MCS combinations attainable with a lead/Lexan sandwich.\label{fig:PbLexan}}
  \end{figure}

For a specific combination of energy loss and MCS angle we can use a `high-$Z$/low-$Z$ sandwich', that is, a combination of Pb and (say) Lexan. Using the techniques already sketched, we can write a program which, given two materials, produces the desired result (\fig{PbLexan}). Here we required 160\,MeV protons to be reduced to 110\,MeV at the same time $\theta_0$ was increased from 0 to 40\,milliradians. There are limits, of course. If we ask for too much energy reduction, the amount of Lexan required will already overscatter (no Pb needed). If we ask for too little energy reduction, no Lexan is needed and the Pb will underscatter.

\begin{figure}[p]
\centering\includegraphics[width=3.26in,height=3.5in]{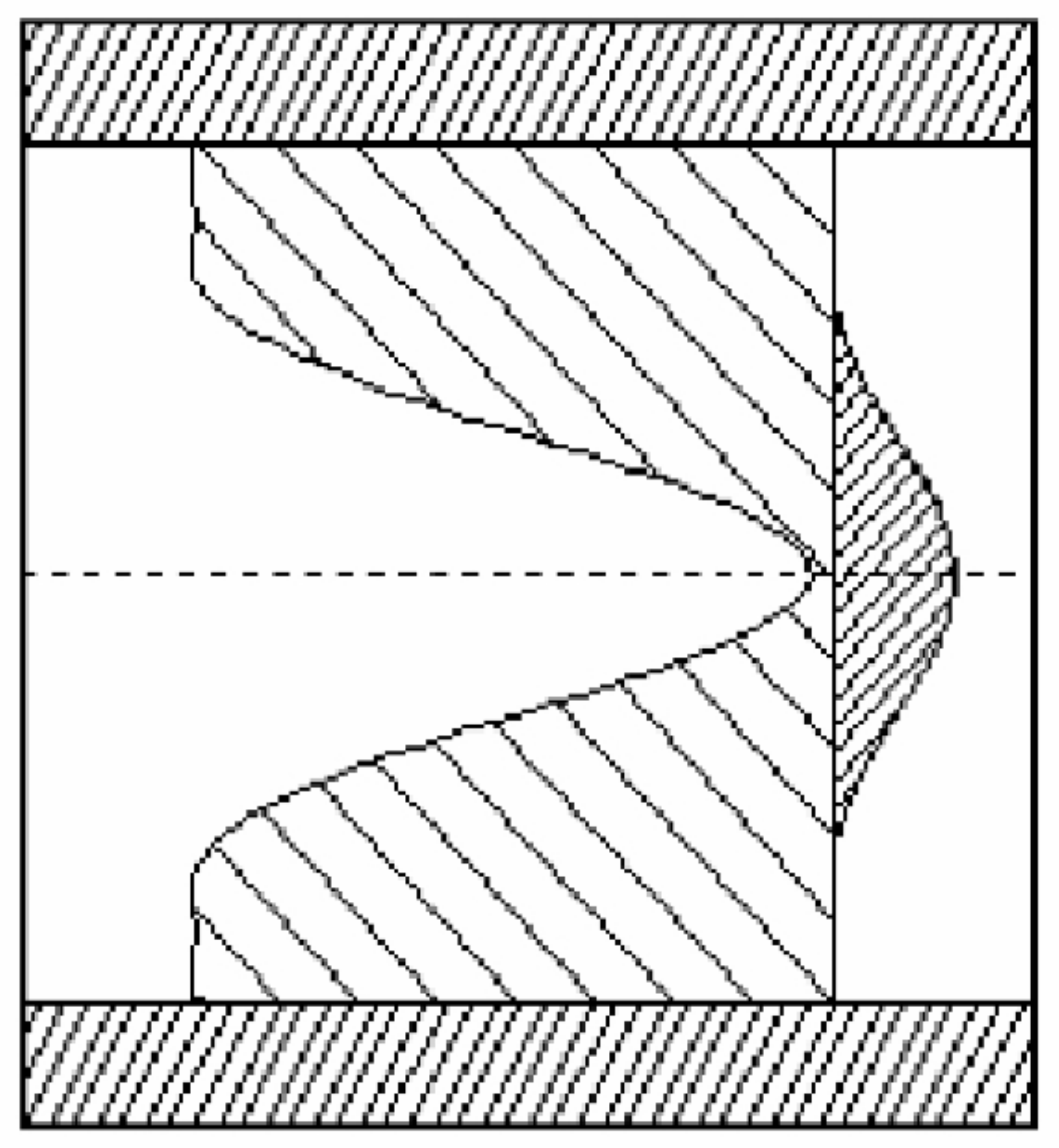}
\caption{A compensated contoured scatterer. Mixing Pb and Lexan yields a desired scattering profile while keeping the energy loss the same at any radius.\label{fig:compScat}}
  \end{figure}

\fig{compScat} shows an application, a compensated contoured scatterer. The combination of the Pb dome (right) and the Lexan (left) produces a desired scattering profile and, at the same time, an energy loss independent of radius.

\subsection{Beam Spreading by Single Scattering}
\begin{figure}[p]
\centering\includegraphics[width=6in,height=3.28in]{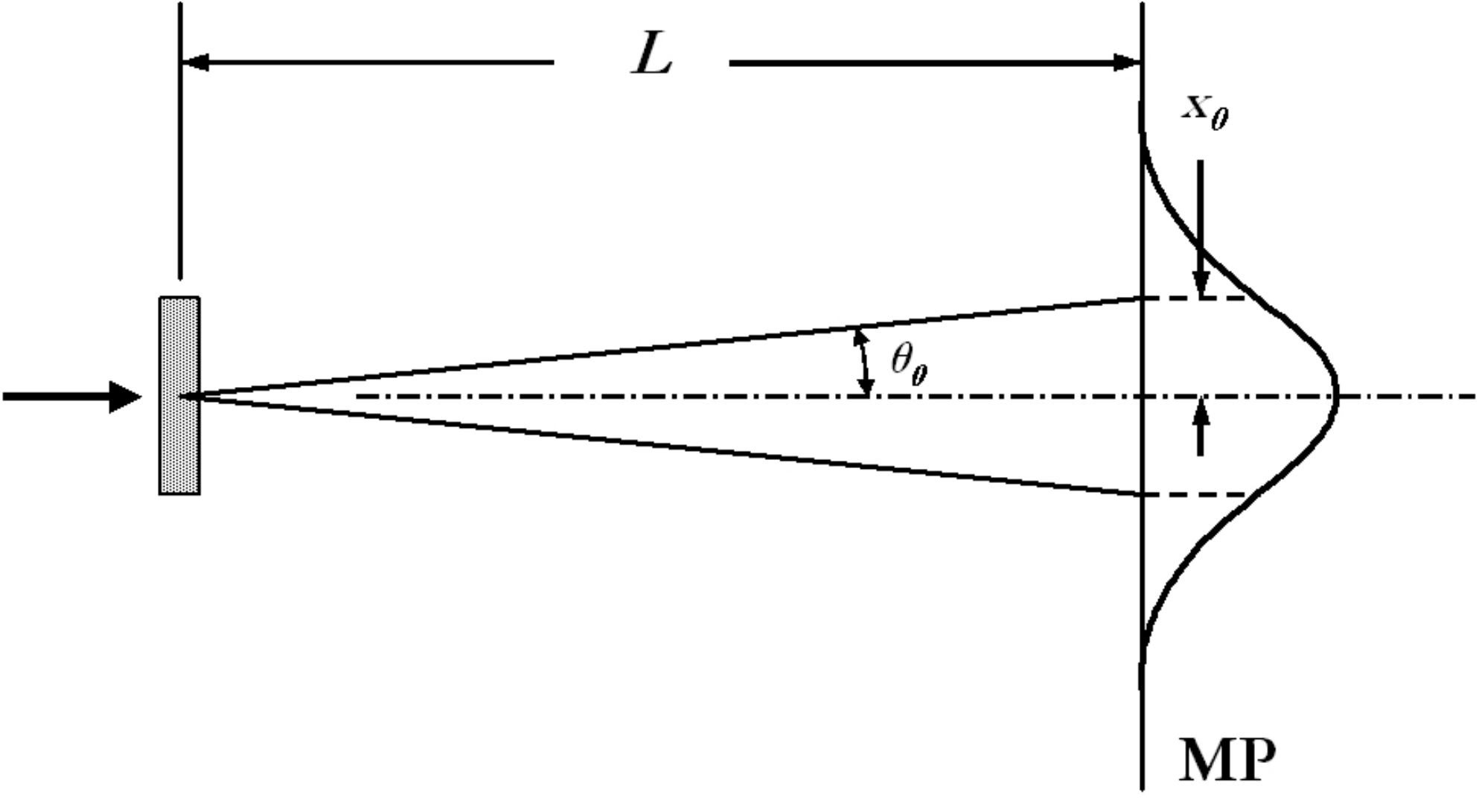}
\caption{Multiple Coulomb scattering in a thin slab.\label{fig:MCSdemo}}
  \end{figure}

\fig{MCSdemo} illustrates a single-scattered therapy beam: a scatterer, a drift or air gap, and an MP at the skin of the patient. A collimator selects the central part of the 2D Gaussian, let us say to a radius $r_{xx}$, so as to get a nearly uniform dose. 

A single scattered beam can be designed by hand, though a proton `desk calculator' such as LOOKUP is very helpful. Let us outline the procedure to cement some of the things we have covered so far. Assume we have a 60\,MeV beam, a 100\,cm throw ($L$) and wish to treat eye lesions (radius 1\,cm) with a dose uniformity of $\pm2.5\%$, using the Gaussian out to the 95\% point ($r_{95}=1$\,cm). From \eqn{G2D}, we require
\begin{equation}
r_0\;=\;r_{95}\;(-2\,\ln0.95)^{-1/2}\;=\;3.122\times r_{95}\;=\;3.122\;\hbox{cm}
\label{eqn:r0}\end{equation}
or an MCS rms angle
\begin{equation*}
\theta_0\;=\;3.122\;\hbox{cm}/100\,\hbox{cm}\;=\;0.0312\;\hbox{rad}
\end{equation*}
We'll use a Pb scatterer to conserve energy. LOOKUP, using the procedure outlined in \sect{MCSprobs}, tells us to use 0.523\,g/cm$^2$ or 0.461\,mm of Pb. Also from LOOKUP, the range of the incident 60\,MeV beam in water is 3.126\,cm, the residual energy after the Pb is 57.3\,MeV and the corresponding residual range is 2.88\,cm H$_2$O, about right for eye treatments.\footnote{~If we want, we can replace the scatterer by a range modulator to cover the extent in depth of the target. Designing that is beyond the scope of this writeup.}

To estimate the dose rate at the MP (proximal surface of the eye) we use \eqn{Di} with $i=1$\,nA, noting from \eqn{G2D} that the fluence per proton at $r=0$ is $1/(2\pi\,r_0^2)$. From LOOKUP, $S/\rho$ for 57.3\,MeV protons in water is $11.1$\,MeV/(g/cm$^2$) so
\begin{equation}\label{eqn:dotD1}
  \dot{D}\;=\;\frac{1\times11.1}{2\pi\times3.122^2}\;=\;0.1812\;\hbox{Gy/s}\;=\;10.9\;\hbox{Gy/min}
\end{equation} 
a bit fast but in the right ballpark. Alternatively, we could have used the treatment area and the efficiency, obtaining
\begin{equation}\label{eqn:dotD2}
  \dot{D}\;=\;\frac{1\times0.05\times11.1}{\pi\times1^2}\;=\;0.1767\;\hbox{Gy/s}\;=\;10.6\;\hbox{Gy/min}
\end{equation}
That is the dose rate averaged over the field, 2.5\% lower (as expected) than the dose rate at $r=0$.

\clearpage
\listoffigures



\clearpage





\clearpage


\clearpage




\clearpage


\clearpage



\clearpage


\end{document}